# Reciprocal spin Hall effects in conductors with strong spin-orbit coupling: a review

Yasuhiro Niimi[1*] and YoshiChika Otani[1,2†]

[1]Institute for Solid State Physics, University of Tokyo, 5-1-5 Kashiwa-no-ha, Kashiwa, Chiba 277-8581, Japan
[2]RIKEN-CEMS, 2-1 Hirosawa, Wako, Saitama 351-0198, Japan

*E-mail: niimi@phys.sci.osaka-u.ac.jp, Present Address: Department of Physics, Osaka University, 1-1 Machikaneyama, Toyonaka, Osaka 560-0043, Japan
†E-mail: yotani@issp.u-tokyo.ac.jp

## Abstract

Spin Hall effect and its inverse provide essential means to convert charge to spin currents and vice versa, which serve as a primary function for spintronic phenomena such as the spin-torque ferromagnetic resonance and the spin Seebeck effect. These effects can oscillate magnetization or detect a thermally generated spin splitting in the chemical potential. Importantly this conversion process occurs via the spin-orbit interaction, and requires neither magnetic materials nor external magnetic fields. However, the spin Hall angle, i.e., the conversion yield between the charge and spin currents, depends severely on the experimental methods. Here we discuss the spin Hall angle and the spin diffusion length for a variety of materials including pure metals such as Pt and Ta, alloys and oxides determined by the spin absorption method in a lateral spin valve structure.

## 1. Introduction

Spintronics is a new class of electronics where two degrees of freedom of an electron, i.e., charge and spin, are utilized [1]. Thus, understanding spin-dependent transport properties is at the heart of spintronics. One of the most important achievements in this field is the discovery of giant magnetoresistance (GMR) observed in thin film structures composed of alternating ferromagnetic and nonmagnetic conductive layers [2, 3]. Depending on the relative alignment of the magnetization in the two ferromagnetic layers (parallel or antiparallel state), a significant change in the electrical resistance can be observed. This type of device is referred to as a spin valve (figure 1(*a*)). Since the nonmagnetic layer is very thin, the spin orientation of conduction electrons is conserved, resulting in GMR. A few years before the discovery of GMR, Johnson and Silsbee [4] succeeded in fabricating a similar spin valve but with a lateral structure (figure 1(*b*)). This device structure is called lateral spin valve. Compared to GMR, the resistance change observed in the lateral spin valve was fairly small, but the important achievement in this work was to generate a pure spin current, that is only a flow of spin angular momentum and does not accompany net charge current. As detailed in section 3, by flowing an electric charge current from a ferromagnetic wire to a nonmagnetic one, spin accumulation is induced at the interface between the two wires. As a result, the pure spin current can flow on the other side of the nonmagnetic wire where no charge currents flow.

The generation of the pure spin current can also be achieved by the spin Hall effect (SHE). The SHE is described as an electric current induced accumulation of spins along the edges of a nonmagnetic conductive wire, and the spin-orbit (SO) interaction in the conductor induces such spin accumulation (figure 2(*a*)). The SHE was originally predicted by Dyakonov and Perel [5] in 1971 and revived by Hirsch [6] about thirty years later. After its experimental observation in GaAs semiconductors in 2004 [7], it attracted a renewed interest for its possible application in spintronics to convert charge into spin currents using neither ferromagnets nor external magnetic fields. However, the spin Hall (SH) angle, characteristic of the conversion yield between charge and spin, was quite small (on the order of $10^{-3}$-$10^{-4}$) in semiconductors [7, 8].

Another breakthrough was the electrical detection of much larger inverse SHE (ISHE) in metallic devices. In the SHE configuration, the converted spin current cannot be directly measured by the electrical method. That is why the SHE has not been observed for more than 100 years after the discoveries of the Hall effect in nonmagnetic metals [9] and the anomalous Hall effect (AHE) in ferromagnetic metals [10]. When the spin current is injected into nonmagnetic materials with strong SO interactions, the charge current can be generated and detected electrically (figure 2(*b*)). In 2006, two important reports on the ISHE in metallic systems were published. One is about the electrical detection of the ISHE using aluminum Hall bar [11]. However the SH angle was still small ($3\times10^{-4}$). At approximately the same time, on the other hand, Saitoh *et al*. [12] observed a large ISHE signal in platinum/permalloy (hereafter Py; $Ni_{81}Fe_{19}$) bilayer structure. So far, the ISHE in Pt has been confirmed by several experimental methods such as spin pumping in a microwave cavity [12, 13], spin pumping with coplanar waveguides [14, 15], spin transfer torque induced ferromagnetic resonance (ST-FMR) [16, 17], SH magnetoresistance [18, 19], and spin absorption in a LSV structure [20, 21]. Thanks to a large SH angle of Pt, it has been utilized as a detector of the spin Seebeck effect which enables to convert heat into spin current [22-27] and also as a generator of spin torque for switching a small ferromagnet [28]. Although Pt is widely believed as one of the best SHE materials, the absolute value of the SH angle (0.01~0.1) is still under debate among several experimental groups [29].

On the other hand, such a large SHE has also been measured in Cu-based alloys [30-33]. The detailed mechanism of the SHE, however, is different from that in 4*d* and 5*d* transition metals such as Pt, Pd, and Ta [34-36]. In the next section, we review both the extrinsic and intrinsic SHEs.

## 2. Spin Hall effects in metallic systems

SHEs in metallic systems are classified into two categories. One is the extrinsic SHE based on the impurity scattering. The other is the intrinsic SHE based on the band structure. These two categories are the same as those in AHEs in ferromagnets, as reviewed in [37]. In the following subsections, we briefly explain both extrinsic and intrinsic SHEs.

*2.1. Extrinsic spin Hall effect*

As in the case of the extrinsic AHE, there are two kinds of mechanisms in the extrinsic SHE, i.e., skew scattering [38] and side jump [39]. The skew scattering is the event which occurs at an impurity site with strong SO interaction. This impurity forms a potential landscape; the resulting scattering bends the trajectories of spin-up and spin-down electrons to different directions (see figure 3(*a*)). During the scattering event, the wave vector is not conserved. In other words, the skew scattering is an inelastic event. On the other hand, the side jump is also the event occurring at the impurity site, but it is an elastic event. As illustrated in figure 3(*b*), at the impurity site there is a discontinuous displacement along the vertical direction for the incident wave vector. Since the displacement is the same for the spin-up and spin-down electrons but with an opposite sign, the total momentum is still conserved.

It is well-known that in the skew scattering the SH resistivity $\rho_{SHE}$ is proportional to the resistivity $\rho_{imp}$ induced by impurities with strong SO interaction [40]. In the side jump, on the other hand, $\rho_{SHE}$ is proportional to $\rho_{imp}^2$ when the impurities are the only source of resistivity, or proportional to $\rho_{imp}\rho_{total}$ when $\rho_{total}$ includes an additional contribution from scattering potentials with weak SO interactions. Here $\rho_{total}$ is the total resistivity. The SH angle $\alpha_H$ is defined as the ratio of $\rho_{SHE}$ and $\rho_{imp}$, i.e., $\alpha_H \equiv \rho_{SHE}/\rho_{imp} = a + b\rho_{imp}$ (or $\rho_{total}$) [41]. Here *a* and *b* are the coefficients of the skew scattering and the side jump. Thus, if there is no impurity with strong SO interaction i.e., $\rho_{imp} = 0$, $\alpha_H$ is also zero even though $\rho_{total}$ is not zero. In the low $\rho_{imp}$ region, the skew scattering is dominant while in the high $\rho_{imp}$ (or $\rho_{total}$) region, the side jump becomes dominant.

*2.2. Intrinsic spin Hall effect*

The intrinsic mechanism was proposed by Karplus and Luttinger in 1954 [42] to explain the AHE, earlier than the above two extrinsic mechanisms (skew scattering [38] and side jump [39]). However, most of the experimental results were explained by the skew scattering and the side jump. Thus, the concept of the intrinsic mechanism had been put aside for a long time. It received a renewed interest after the Karplus and Luttinger theory was reformulated by the Berry phase language [43-45]. In fact, it can quantitatively explain the observed AHEs not only in pure ferromagnetic metals but also in ferromagnetic semiconductors and oxides [46].

The same mechanism can be applied for the SHEs in 4*d* and 5*d* transition metals such as Pt, Pd and Ta. According to a recent spin pumping measurement using a ferromagnetic insulator, yttrium iron garnet YIG [47], the intrinsic mechanism is also important for the generation of SHEs in $3d$ transition metals such as V and Ni. As detailed in [34] and [35], the degeneracy of $d$-orbitals subjected to the LS coupling plays an important role for generating the intrinsic SHE. In the intrinsic mechanism, the SH resistivity $\rho_{SHE}$ is proportional to $\rho_{xx}^2$ when the resistivity of transition metal $\rho_{xx}$ is relatively small. On the other hand, when $\rho_{xx}$ is larger than some critical value, $\rho_{SHE}$ rapidly decreases [35].

## 3. Spin Hall angle and spin diffusion length

To characterize the SHEs in nonmagnetic materials, there are two important parameters, i.e., SH angle and spin diffusion length. There is a heavy debate among several groups engaged in spintronics about how to determine the two parameters correctly [29]. In this section, we first explain the spin absorption method in the lateral spin valve structure to obtain the SH angle and spin diffusion length. As a complementary method to estimate the spin diffusion length in nonmagnetic metals, weak antilocalization can be used, as detailed in section 3.4.

*3.1. Spin absorption method with lateral spin valve structure*

The lateral spin valve is one of the methods to generate a pure spin current, as briefly mentioned in section 1 and figure 1(*b*). When an electric charge current $I_C$ is passed from one of the ferromagnetic wires (F; injector) to a nonmagnetic wire (N) with weak SO interaction (figure 4(*a*)), spin accumulation ($\delta\mu$) is generated at the interface between F and N so that the spin-up and spin-down chemical potentials are continuous at the interface (figures 4(*b*) and 4(*c*)). Thereby, only a pure spin current $I_S$ flows diffusively on the other side of N wire, and decays over a characteristic length, so-called spin diffusion length $\lambda_N$ (figure 4(*b*)). If another ferromagnetic wire (F; detector) is attached to the N wire within $\lambda_N$, a finite nonlocal voltage $V_S$ is generated and changes depending on the magnetization of the two ferromagnets, as illustrated in figure 4(*c*). This measurement configuration is called nonlocal spin valve (NLSV). The voltage drop $\Delta V_S$ between the two states is proportional to the spin accumulation at the detector position. By changing the distance $L$ between the two F wires, $\lambda_N$ can be estimated. This has already been formulated by Takahashi and Maekawa within the one-dimensional (1D) spin diffusion model [48], based on the Valet-Fert model for GMR configuration [49], and experimentally studied first by Jedema *et al* [50-52]. However this analysis can only be applicable to N with $\lambda_N$ longer than approximately 100 nm because of the technical limitation, i.e. the fabricable size by the electron beam lithography. In order to overcome this technical limitation, the spin absorption method is very powerful to determine the spin diffusion length of metal with strong SO interaction (< ~100 nm).

Figure 5(*a*) shows the principle of the spin absorption method. It is based on the lateral spin valve device, and a target metal (M) with strong SO interaction is placed just in the middle between the two F wires. Since the SO interaction of M is much stronger than that of N, most of $I_S$ are absorbed (flows) perpendicularly into the M wire. When the magnetization is aligned along the hard direction of the injector, conduction electron spins also point to the same direction. As illustrated in figure 5(*a*), since the directions of $I_S$ and spin angular momentum $s$ of conduction electrons are orthogonal to each other, $I_S$ is converted into $I_C$ via the ISHE along the M wire direction. The relation-

ship among the three directions is given by $I_C \propto I_S \times s$.

As a matter of fact, all the $I_S$ are not absorbed into the M wire but a small amount of $I_S$ remain flowing along the N channel. The remaining pure spin current can be detected by the detector, as in the case of the lateral spin valve device without M (figure 5(*b*)). As detailed in the following subsections, by comparing the spin accumulation with and without the M wire, we can estimate $I_S$ absorbed into M and also the spin diffusion length of M, $\lambda_M$.

It is noteworthy that the direct SHE (DSHE) can also be measured on the same device just by swapping the voltage probes for the current leads. By flowing $I_C$ along the M wire, the spin accumulation is generated at the interface between M and N. This spin accumulation can be detected by the F wire [20, 21].

*3.2. One-dimensional spin diffusion model*

To quantitatively evaluate the SH angle $\alpha_H$ and the spin diffusion length $\lambda_M$ of a SH material in the lateral spin valve device, one needs to extend the Takahashi and Maekawa model to the case of 3-wire configuration. In the early stage of SHE measurements with the spin absorption method [20, 21], two important points were overlooked. Firstly the equations to extract $\alpha_H$ and $\lambda_M$ were not correctly derived. The second point is more problematic than the first one; that is the shunting effect due to the N channel above the M wire. Especially, this caused underestimation of $\alpha_H$ by a factor of about 3-4. In the following, we will explain the 1D spin diffusion model.

In the spin absorption method, the SH angle $\alpha_H$ is obtained from the ISHE voltage $\Delta V_{ISHE}$:

$$\alpha_H \equiv \frac{\rho_{SHE}}{\rho_{xx}} = \frac{w_M}{x\rho_{xx}} \frac{\Delta V_{ISHE}}{\bar{I}_S} \quad (1).$$

Note that for Cu-based SHE materials, $\rho_{xx}$ has to be replaced by $\rho_{imp}$ ($=\rho_{xx} - \rho_{Cu}$) since Cu itself does not contribute to the SHE as detailed in [40]. $w_M$ is the width of the M wire and $x$ is the shunting factor as we will describe later on. $\bar{I}_S$ is the effective pure spin current injected vertically into the M wire:

$$\bar{I}_S = \frac{\int_0^{t_M} I_S(z)dz}{t_M} = \frac{\lambda_M}{t_M} \frac{\{1-\exp(-t_M/\lambda_M)\}^2}{1-\exp(-2t_M/\lambda_M)} \frac{2p_F I_C Q_F \{\sinh(L/2\lambda_N)+Q_F \exp(L/2\lambda_N)\}}{\{\cosh(L/\lambda_N)-1\}+2Q_M \sinh(L/\lambda_N)+2Q_F \{\exp(L/\lambda_N)(1+Q_F)(1+2Q_M)-1\}}$$

(2),

$t_M$ is the thickness of the M wire. $Q_F$ and $Q_M$ are defined as $R_F/R_N$, and $R_M/R_N$, where $R_F$, $R_N$, and $R_M$ are the spin resistances of the F, N, and M wires, respectively. The spin resistance $R_X$ (X = F, N, or M) is different from the normal resistance and expressed as $R_X = \rho_X \lambda_X / \{(1-p_X^2)A_X\}$. Here $\rho_X$, $\lambda_X$, and $p_X$ are the resistivity, the spin diffusion length, and the spin polarization of material X. Since N and M are nonmagnetic metals, $p_N = p_M = 0$. $A_X$ is the effective cross section for pure spin current in material X. $A_N = w_N t_N$ for $\lambda_N \gg t_N$, $A_F = w_N w_F$ for $\lambda_F \ll t_F$ and $A_M = w_N w_M \tanh(t_M/\lambda_M)$ for $t_M$, $\lambda_M \ll w_N$, where $w_N$, $w_F$, and $t_N$ are the widths of the N and F wires and the thickness of the N channel, respectively. The perpendicularly absorbed pure spin current in the M wire attenuates exponentially when $\lambda_M < t_M$, but linearly down to zero at the bottom surface of the M wire when $\lambda_M > t_M$. Since $\lambda_N$, $\lambda_F$, and $p_F$ can be obtained from the *L* dependence of NLSV signals as mentioned in the previous subsection (or see supplemental material in [31] for more details), only $\lambda_M$ is the remaining physical quantity to obtain $\bar{I}_S$.

The spin diffusion length $\lambda_M$ of the M wire can be determined from NLSV measurements. The ratio $\eta$ of the NLSV voltages with and without the M wire is expressed as:

$$\eta \equiv \frac{\Delta V_S^{with}}{\Delta V_S^{without}} = \frac{2Q_M \{\sinh(L/\lambda_N) + 2Q_F \exp(L/\lambda_N) + 2Q_F^2 \exp(L/\lambda_N)\}}{\{\cosh(L/\lambda_N) - 1\} + 2Q_M \sinh(L/\lambda_N) + 2Q_F \{\exp(L/\lambda_N)(1+Q_F)(1+2Q_M) - 1\}} \quad (3).$$

By substituting all the values in equation (3), $\lambda_M$ as well as $Q_M$ can be evaluated.

The coefficient $x$ in equation (1) is so-called the shunting factor. This factor expresses the magnitude of shunting by the Cu contact above the M material. Its value, $x \approx 0.36$, can be found by comparing the resistance of the M wire to that of the M wire with a 100 nm wide Cu bridge, as described in [30]. However, there was a controversial issue on how to evaluate the shunting factor $x$ [29]. The evaluation of $x$ is very crucial to determine the SH angle correctly. In fact, the shunting effect is automatically taken into account in a three-dimensional (3D) finite element analysis, as described in the next subsection.

*3.3. Three-dimensional spin diffusion model*

In spite of taking into account the shunting factor $x$, there was still a criticism that $x$ was too large to explain large $\alpha_H$ obtained with the spin pumping and ST-FMR measurements with ferromagnet/Pt or ferromagnet/Ta bilayer films. To overcome the criticism, we have employed a 3D spin diffusion model based on the 1D Valet-Fert equation [49] to evaluate the SH angle and the spin diffusion length.

The 1D Valet-Fert theory can be extended to 3D and arbitrary noncollinear magnetization configurations. In the Valet-Fert theory, there are two fundamental equations; charge and spin drift-diffusion equations. When considering the possibility of SHE in nonmagnetic material (M) with strong SO interaction, appropriate new off-diagonal conductivity terms should be added to the two equations (these off-diagonal terms are neglected in nonmagnetic material (N) with very weak SO interaction). In tensorial format (a summation is implied for the repeated indices), the new equations read:

$$J^\alpha = -\frac{\nabla_\alpha \mu}{e\rho_M} + \alpha_H \varepsilon_{\alpha\beta\gamma} \frac{\nabla_\beta \mu_S^\gamma}{2e\rho_M} \quad (4),$$

$$J_S^{\alpha,\beta} = -\alpha_H \varepsilon_{\alpha\beta\gamma} \frac{\nabla_\gamma \mu}{e\rho_M} - \frac{\nabla_\alpha \mu_S^\beta}{2e\rho_M} \quad (5),$$

where $e$, $J$ and $\mu$ are, respectively, the electric charge, the charge current density, and the electro-chemical potential. The index S indicates a spin quantity and $\varepsilon_{\alpha\beta\gamma}$ is the Levi Civita symbol in 3D. Using the same notation, the charge and spin drift-diffusion equations in the F wire can be expressed as:

$$J^\alpha = -\frac{\nabla_\alpha \mu}{e\rho_F} - \frac{\beta \nabla_\alpha \mu_S}{2e\rho_F} \quad (6),$$

$$J_S^\alpha = -\frac{\beta \nabla_\alpha \mu}{e\rho_F} - \frac{\nabla_\alpha \mu_S}{2e\rho_F} \quad (7),$$

where $\beta$ is the spin polarization. We assume that the spin quantization axis is parallel to the local magnetization vector in F because of its short transverse spin decoherence length [53-55]. As we will see in the next section, $\beta$ is slightly different from $p_F$ obtained from the 1D Takahashi and Maekawa formula.

Numerical calculations based on the 3D version of the Valet-Fert model have been performed using SpinFlow 3D [31, 33]. It implements a finite element method to solve a discrete formulation of the bulk transport equations, supplemented with the interface and boundary conditions. In the spin absorption type lateral spin valve, there are two interfaces; F and N, and N and M. The interface between F and N is characterized by three parameters: the interfacial areal resistance $r_b^*$, the interfacial resistance asymmetry coefficient $\gamma$, and the spin mixing conductance $g_{\uparrow\downarrow}$. As for $r_b^*$ and $g_{\uparrow\downarrow}$, we take their values from appropriate references [56-58], while $\gamma$ as well as $\beta$ can be determined by analyzing the NLSV signal without the M wire as a function of $L$, as in the case of the 1D model. On the other hand, the interface between N and M is simply characterized by $r_b^*$. It should be noted that the shunting effect between N and M is automatically taken into account in this 3D finite element calculation.

*3.4. Weak antilocalization as alternative method to determine spin diffusion length*

As described above, the SH angle and the spin diffusion length of a SH material can be evaluated using the spin absorption method in the lateral spin valve structure on the same device. However, there has been a heavy debate about the validity of the SH angle and the spin diffusion length deduced by this method [29]. Especially, the spin diffusion length of Pt determined by the spin absorption method [33, 36, 55] is often several times longer than that obtained with nonmagnet/ferromagnet bilayer films [16-19, 59]. To judge whether the spin diffusion length obtained with the spin absorption method is too large or not, another approach is highly desirable.

Weak antilocalization (WAL) is one of the simple ways to obtain the spin diffusion length, as already reported in previous papers [33, 55, 60, 61]. Weak localization occurs in metallic systems and has been used to study decoherence of electrons [62-65]. The principle of this technique relies on constructive interference of closed electron trajectories which are traveled in opposite direction (time reversed paths). This leads to an enhancement of the resistance. The magnetic field $B$ perpendicular to the plane destroys these constructive interferences, leading to a negative magnetoresistance $R(B)$ whose amplitude and width are directly related to the phase coherence length. If there is a non-negligible SO interaction, a positive magnetoresistance can be obtained, which is referred to as WAL [66].

The dimension of the system is determined with respect to the phase coherence length $L_\varphi$ and the elastic mean free path $l_e$. Since we deal with nanometer-scale metallic systems, $l_e$ is in general smaller than all the sample dimensions. On the other hand, the inelastic scattering length $L_\varphi$ can be relatively long for a clean metallic system. When $L_\varphi$ is larger than the width $w$ and the thickness $t$ of sample but smaller than the length $l$, we call the system "quasi-1D".

The WAL peak of quasi-1D wire can be fitted by the Hikami-Larkin-Nagaoka formula [62, 66]:

$$\frac{\Delta R}{R_\infty} = \frac{2}{l} \frac{R_\infty}{h/e^2} \left( \frac{\frac{3}{2}}{\sqrt{\frac{1}{L_\varphi^2} + \frac{4}{3}\frac{1}{L_{SO}^2} + \frac{1}{3}\frac{w^2}{l_B^4}}} - \frac{\frac{1}{2}}{\sqrt{\frac{1}{L_\varphi^2} + \frac{1}{3}\frac{w^2}{l_B^4}}} \right) \quad (8),$$

where $\Delta R$, $R_\infty$ and $L_{SO}$ are the WAL correction factor, the resistance of the wire at high enough field, and the SO length, respectively. $h$ and $l_B \equiv \sqrt{(h/2\pi eB)}$ are the Plank constant and the magnetic length, respectively. In equation (8), we have only two unknown parameters; $L_\varphi$ and $L_{SO}$. According to the Fermi liquid theory [64, 65, 67], $L_\varphi$ depends on temperature ($\propto T^{-1/3}$), while $L_{SO}$ is almost constant at low temperatures [63].

The relation between the SO length and the spin diffusion length has been theoretically discussed [68] and the schematics of the two length scales are depicted in figure 6. The SO scattering rate $1/\tau_{SO} = D/L_{SO}^2$ includes both spin-flip and spin-conserving processes. Here $D$ is the diffusion constant. Thus one obtains $1/\tau_{SO} = 3/(2\tau_{\uparrow\downarrow})$ where $1/\tau_{\uparrow\downarrow}$ is the spin-flip scattering rate. We also note that the spin relaxation rate $1/\tau_s = D/L_s^2$ is twice the spin-flip scattering rate, i.e., $1/\tau_s = 1/\tau_{\uparrow\downarrow} + 1/\tau_{\downarrow\uparrow}$. At sufficiently low temperatures, we can neglect the contribution of phonons and obtain the following relation:

$$L_s = \frac{\sqrt{3}}{2} L_{SO} \quad (9),$$

within the Elliott-Yafet mechanism [69, 70] from isotropic impurity scattering. As can be seen in [55], equation (9) has been verified experimentally. Since $L_s$ is basically equivalent to $\lambda_N$ or $\lambda_M$, we use hereafter only $\lambda_N$ or $\lambda_M$ as the spin diffusion length of nonmagnetic metal.

## 4. Experimental details

In this section, we show some experimental details about how to prepare SH devices and WAL samples and also how to measure the SHE and WAL.

### 4.1. Sample preparations

As shown in figure 5, the SH device consists of two F wires and one M wire which are bridged by a thick N wire. In the present study, we have used Py as F and Cu as N. The wires were prepared using electron beam lithography onto a thermally oxidized silicon substrate coated with polymethyl-methacrylate (PMMA) resist or ZEP 520A.

A pair of Py wires was first deposited using an electron beam evaporator under a base pressure of $10^{-9}$ Torr. The thickness $t_F$ and width $w_F$ of the Py wires are 30 nm and 100 nm, respectively. The M wire was next deposited by 20 nm ($t_M$ = 20 nm) by sputtering the M target or by heating it with an electron beam evaporator. Especially, the sputtering was used for Cu-based and Ag-based alloys or for high-melting-point metals such as Ta and Nb. In table I, we show the list of the M wires presented in this paper [30, 31, 33, 36]. After preparing the Py and M wires, a 100 nm wide and 100 nm thick Cu wire was deposited on the three wires by a Joule heating evaporator using a 99.9999% purity source. Before deposition of the Cu bridge, we performed a careful Ar ion beam etching for 30 seconds in order to clean the surfaces of Py and the M middle wires. Concerning WAL samples, we prepared ~1 mm long, 100-120 nm wide, and 20 nm thick (except for figures 12(*a*) and 12(*b*)) nonmagnetic metal wires to obtain larger $\Delta R$ signals and also to meet the condition that equation (8) is valid. Typical scanning electron micrographs of the SH device and the WAL sample are shown in figures 7(*a*) and 7(*b*), respectively.

*4.2. Measurement methods*

Both SHE and WAL measurements have been carried out using an ac lock-in amplifier (modulation frequency $f$ =173 Hz) and a $^4$He flow cryostat. In order to obtain a very small WAL signal compared to the background resistance, we used a bridge circuit as detailed in [65]. To check the reproducibility and to evaluate the error bars, we have measured 3-6 different samples.

To determine the spin polarization of Py as well as the spin diffusion lengths of Py ($\lambda_F$) and Cu ($\lambda_N$), the NLSV measurements without the M wire have already been performed. Both 1D and 3D models give the same $\lambda_F$ and $\lambda_N$, but the spin polarization is slightly different. For example, $\beta = \gamma = $ 0.31 at 10 K obtained from the 3D Valet-Fert model, while $p_F$ = 0.23 at 10 K obtained from the 1D Takahashi-Maekawa model [71]. We summarize the detailed values of physical quantities related to the NLSV measurements in table II.

## 5. Extrinsic spin Hall effects in alloys

*5.1. Extrinsic spin Hall effects in Cu -based alloys*

Using the SH device shown in figure 7(*a*), both DSHE and ISHE can be measured. Figures 8(*a*) and 8(*b*) show typical DSHE and ISHE signals $R_{D(I)SHE}$ (D(I)SHE voltages $V_{D(I)SHE}$ divided by an electric current $I_C$ from Py to Cu) of Ir (3 atomic percent)-doped Cu, Bi (0.5 atomic percent)-doped Cu, and pure Cu. Note that the magnetic field is applied to the hard direction of the Py wire such that the condition $I_C \propto I_S \times s$ can be fulfilled. The SHE of Cu is negligibly small, but when Ir or Bi impurities are added in Cu, the alloys show SHE signals. Both $R_{DSHE}$ and $R_{ISHE}$ linearly increases with the magnetic field and saturate above 2000 Oe, corresponding to the saturation field of the magnetization, as can be seen in the anisotropic magnetoresistance (AMR) curve of figure 8(*c*). The amplitude of the SHE resistance $\Delta R_{SHE}$ is exactly the same for both the DSHE and ISHE, in agreement with the Onsager reciprocal relation [20, 21, 30, 31]. When we compare $Cu_{99.5}Bi_{0.5}$ and $Cu_{97}Ir_3$, the former has a several times larger SHE signal than the latter in spite of a smaller doping. In addition, the sign of the SHE in CuBi (negative) [31] is opposite to that of CuIr (positive) [30].

In order to evaluate the SH angles $\alpha_H$ of CuIr and CuBi alloys from the above SHE measurements, one needs to obtain the spin diffusion lengths $\lambda_M$ of these alloys. Thus, we next see the spin absorption effect by inserting the CuIr and CuBi alloys from NLSV measurements. For the NLSV measurements, the magnetic field is applied along the Py wire direction (see figure 5(*b*)). Typical NLSV signals $R_S$ (nonlocal voltages $V_S$ divided by $I_C$) with and without the M ($Cu_{97}Ir_3$ or $Cu_{99.5}Bi_{0.5}$) wire is shown in figure 9. Apparently, by inserting the M wire, the NLSV signal is reduced. From the reduction in the NLSV signal, the spin diffusion length $\lambda_M$ of the M wire can be extracted using the 1D and 3D spin diffusion models as already detailed in section 3. In figure 10(*a*) and 10(*b*), we show $\lambda_M$ of the CuIr and CuBi alloys with different impurity concentrations, obtained from the 1D and 3D models. In general, $\lambda_M$ exponentially decreases with increasing the concentration of Ir or Bi impurities. In the case of CuIr, the difference between $\lambda_M^{1D}$ and $\lambda_M^{3D}$ is relatively small. On the other hand, the difference between the two is not negligible for lower concentrations of Bi in Cu. This difference originates from the spreading of spin accumulation along the side arms of the CuBi/Cu cross junction.

Figures 11(*c*) and 11(*d*) show the 3D mappings of spin accumulation for $Cu_{99.5}Bi_{0.5}$ and $Cu_{97}Ir_3$ devices, respectively. For the $Cu_{97}Ir_3$ device, the spreading of spin accumulation is not so obvious, but it is highly visible for the $Cu_{99.5}Bi_{0.5}$ device. The spin diffusion length $\lambda_M$ of $Cu_{99.5}Bi_{0.5}$ is longer than the thickness $t_M$. In this situation, a part of pure spin current is not absorbed vertically into the $Cu_{99.5}Bi_{0.5}$ wire but spreads along the wire. Since we assume in the 1D model that all the pure spin current is absorbed vertically into the M wire, such an effect of the spreading is not taken into account in the 1D model, which leads to underestimation of $\lambda_M$ as well as $\alpha_H$. On the other hand, if the condition $\lambda_M < t_M$ is fulfilled like in the case of CuIr, the 1D model explains the spin diffusion length fairly well.

Once the spin diffusion length of the M wire is obtained, the SH angle $\alpha_H$ of the M wire can be evaluated using equations (1) and (2) for the 1D model and equations (4) and (5) for the 3D model. The SH resistivities $\rho_{SHE}$ of the CuBi and CuIr alloys obtained from the 1D (open symbols) and 3D models (closed symbols) are shown in figures 11(*a*) and 11(*b*), respectively. Note that the horizontal axis is the resistivity induced by the Ir or Bi impurity, i.e., $\rho_{imp}$. The SH resistivity $\rho_{SHE}$ of CuIr increases linearly with increasing the concentration of Ir impurities up to 12%. This linear dependence is a clear evidence of the skew scattering in the CuIr alloys, as explained in section 2.1. We have also checked the temperature dependence of the SH angle. It is constant with temperature, which is also typical of the skew scattering [30]. The difference between $\rho_{SHE}^{1D}$ and $\rho_{SHE}^{3D}$ is quite small, which is similar to the case of the spin diffusion length. The slope of the $\rho_{SHE}$ vs $\rho_{imp}$ curve gives the SH angle of CuIr $\alpha_H^{1D} = 0.021$ and $\alpha_H^{3D} = 0.023$, that are quantitatively consistent with the value obtained from previous AHE measurements of CuMnIr ternary alloys [40, 72].

In the case of the CuBi alloys, on the other hand, $\rho_{SHE}$ does not follow the linear law in the whole resistivity region and the linear region is limited up to Bi concentration of 0.5%. This is due to the limited solubility of Bi in Cu [31, 73]. In addition, $\rho_{SHE}^{3D}$ is about two times larger than $\rho_{SHE}^{1D}$ in the linear region. In the low impurity concentration regime, the spin diffusion length is longer than the thickness. Thus, as discussed above, the 1D model underestimates the spin Hall angle and the spin diffusion length. From the slope of the $\rho_{SHE}$ vs $\rho_{imp}$ curve in the linear regime, the SH angle of CuBi is evaluated to be $\alpha_H^{1D} = -0.12$ and $\alpha_H^{3D} = -0.24$. These values are much larger than $\alpha_H$ of the CuIr alloys. Such a large SH angle was first predicted by Grandhand *et al.* [74] but the predicted sign of the SH angle is opposite to the present result. This sign problem has been discussed from the theoretical viewpoints in [75-78]. Although we will not discuss the details, the SH angle should be defined as $\rho_{yx}/\rho_{xx}$ or $\sigma_{xy}/\sigma_{xx}$ (not $\sigma_{yx}/\sigma_{xx}$) to compare with the experimental results and thus all the theoretical calculations predict the negative SH angle for CuBi [74-77]. On the other hand, to reproduce the positive SH angle of CuIr, the on-site Coulomb interaction has to be taken into account [78].

We have also performed similar SHE measurements with other host and impurity metals [36]. For example, when another 6*p* metal Pb is implanted in Cu, a large SH angle is observed as listed in table I. As detailed in [36], however, the SHE signal in $Cu_{99.5}Pb_{0.5}$ disappears in a few days because of the fast diffusion of Pb in Cu [79]. On the other hand, such a time evolution of the SHE has never been observed in CuBi alloys. We have also changed the host metal from Cu to Ag. The SH angles of AgIr and AgBi are not as large as those of CuIr and CuBi [30, 36].

*5.2. Other methods to evaluate spin Hall angle and spin diffusion length*

As described above, the spin absorption in the lateral spin valve structure is a powerful method to evaluate both the SH angle and the spin diffusion length of nonmagnetic metal with strong SO interaction on the same device. Nevertheless, there has been a heated discussion on the SH angle and the spin diffusion length of 4$d$ and 5$d$ transition metals such as Pt [29]. Thus, some other essentially different methods have to be utilized to justify the spin absorption method. Among them, we focus on methods without using ferromagnet/nonmagnet bilayer films, since in such bilayer films the proximity effect between F and N is sometimes non-negligible for the evaluations of the spin diffusion length and the SH angle [80-82]. Thus, we have used WAL measurements [55] for the evaluation of the spin diffusion length, and AHE measurements of CuMnX (X: impurity metal) ternary alloys [40] for the evaluation of the SH angle. The former is already detailed in section 3.4 and the latter is also a well-studied method.

We first tested equation (9). Figure 12($a$) shows WAL curves of pure Cu wires with different thicknesses measured at $T = 4$ K. Clear positive magnetoresistance is observed, which is characteristic of WAL. By fitting the WAL curves with equation (8), $L_{SO}$ can be obtained and converted into the spin diffusion length using equation (9). Figure 12($b$) shows the spin diffusion length of Cu as a function of the diffusion constant $D$. For example, the spin diffusion length of 100 nm Cu wires obtained from equation (9) is about 1000 nm, which is quantitatively consistent with the value obtained from the lateral spin valve [71]. This fact clearly shows that the spin diffusion length of nonmagnetic metal can also be evaluated from the WAL measurement. When we decrease the thickness of Cu, its spin diffusion length also decreases. As can be seen in the inset of figure 12($b$), the elastic mean free path $l_e$ is limited by the thickness since the surface of the Cu wire works as a scatterer. Thus the spin diffusion length of Cu decreases linearly with decreasing $D$. This linear dependence is a clear evidence of Elliot-Yafet mechanism in our system.

Next we show a typical WAL curve of Cu$_{99.5}$Bi$_{0.5}$ wire measured at $T = 4$ K in figure 12 ($c$). Using the same method as shown above, the spin diffusion length of Cu$_{99.5}$Bi$_{0.5}$ can be obtained by using equation (9). It is again quantitatively consistent with $\lambda_M^{3D}$ from the spin absorption measurement. These results verify that the WAL method is quite useful to obtain the spin diffusion length of nonmagnetic metal not only with a weak SO interaction such as Cu and Ag, but also with a strong SO interaction such as CuBi alloys.

We now discuss a different method to obtain the SH angle of Cu-based alloy, i.e., AHE measurement with CuMnX ternary alloy. This was already demonstrated more than three decades ago [40]. As for CuIr alloys, the SH angle obtained from the AHE is quantitatively consistent with that from the SH device. Thus, the AHE measurement has been performed also for our CuMnIr and CuMnBi alloys. The estimated SH angles of Ir-doped Cu and Bi-doped Cu obtained from the AHE measurements are 0.018 and −0.23, respectively. These are also quantitatively consistent with the values with the SH devices of CuIr and CuBi alloys.

## 6. Intrinsic spin Hall effects in 4$d$ and 5$d$ transition metals

The studies on extrinsic SHEs in CuIr and CuBi alloys reveal that the spin absorption in the lateral spin valve structure is a reliable method to evaluate the SH angle and the spin diffusion length. Using the same device structure, we have measured SHEs in 4$d$ and 5$d$ transition metals such as Pt, Pd, and Ta. The SHEs in 4$d$ and 5$d$ transition metals are frequently utilized in modern spintronic experiments, for example for detection of spin Seebeck effect [22-27, 83, 84], efficient magnetization switching [28, 85] and domain wall motion [86]. From these studies, the SH angles of 4$d$ and 5$d$ transition metals are evaluated but they are often much larger than the values estimated from the spin absorption method. In these studies, the 4$d$ and 5$d$ transition metals are always put on or under ferromagnets, which may induce magnetic moments at the interface even in the transition metals due to the proximity effect. On the other hand, it is well-known that such a proximity effect between Cu and ferromagnet is quite weak [80]. Thus, in order to extract the effect only from the SHE and to evaluate the SH angle and the spin diffusion length correctly, it is better to avoid using the ferromagnet/nonmagnet bilayer structure.

Figure 13($a$) shows $R_{\text{ISHE}}$ of several 4$d$ and 5$d$ transition metals. The sign of the SHE depends on the number of $d$-electrons. When 4$d$ or 5$d$ orbital is occupied by electrons less (more) than 5, the sign of the SHE is negative (positive). To obtain the spin diffusion length $\lambda_\text{M}$ and also $I_\text{S}$ injected into the M wire, the NLSV signal with the M wire is compared with the reference NLSV signal in figure 13($b$). From the absorption rate $\eta$, $\lambda_\text{M}$ is evaluated as listed in table I. Except for Au, the spin diffusion lengths of 4$d$ and 5$d$ transition metals are rather short (less than ~10 nm). To double-check the spin diffusion lengths of Pt and Au, WAL measurements were performed at low temperatures as shown in figures 13($c$) and 13($d$), respectively. The spin diffusion lengths obtained from the WAL curves are also consistent with $\lambda_\text{M}$ estimated from the spin absorption measurements (see table I).

As we have described in section 5, the SH angles of the 4$d$ and 5$d$ transition metals were estimated and summarized in figure 14. For comparison, the SH angles theoretically calculated with the model in [35] are also shown in figure 14. The experimental results are in good agreement with the theoretical ones. The SH angle changes the sign at the number of $s+d$ electrons of 7 or 8. We also note that the SH angle of Au reaches half of that of Pt and is larger than Ta, while the raw SHE signal is much smaller than those of Pt and Ta. This is related to the fact that the resistivity of Au is also small and thus the spin diffusion length is relatively long. Thus, we make a point that the evaluation of the SH angle is not simple at all but rather complex since the resistivity, spin diffusion length, shunting effect (in the case of the present SH device), and possible proximity effect at the interface (for F and N bilayer structures) are entangled each other. It is important to obtain all the parameters on the same device, and careful evaluations of these parameters are definitely needed for quantitative discussions on the SH angle.

Before closing this section, let us mention recent progress in the effect of spin memory loss on the spin pumping experiment. The spin memory loss was intensively studied in current-perpendicular-to-plane GMR structures [60, 61, 87-91], in order to explain an additional spin-flip process at interfaces between two different materials. Recently, Rojas-Sánchez *et al.* [92] have included such an effect and estimated the partial depolarization of the spin current at interfaces of Co/Pt bilayers and Co/Cu/Pt trilayers (in the latter case, Cu is inserted to suppress possible induced-magnetic moments of Pt). As a result, the spin diffusion length and the SH angle of Pt are 3.4 nm and 0.056 at room temperature, respectively. These values become little bit closer to the ones determined with the spin absorption method.

We have also considered the spin memory loss in the spin absorption experiment. We suppose a fictitious interfacial layer with the thickness of $t_\text{I}$, the spin diffusion length of $\lambda_\text{I}$, and the resistivity of $\rho_\text{I}$ between N and M. The spin resistance of M is changed from $R_\text{M}$ without the interfacial layer to

$$R_\text{M}' = R_\text{M} \frac{1 - \frac{R_\text{I}}{R_\text{M}} \tanh(\delta)}{1 - \frac{R_\text{M}}{R_\text{I}} \tanh(\delta)} \quad (10),$$

where $\delta \equiv t_I/\lambda_I$ is the spin memory loss parameter introduced in the previous studies [60, 61, 87-92] and $R_I \equiv \rho_I \lambda_I / w_M w_N$ is the spin resistance of the interfacial layer. Interestingly, when $R_M = R_I$, the real spin diffusion length as well as the real spin resistance $R_M'$ of M are independent of the spin memory loss parameter. For our Pt wire, for example, $R_M \sim 0.05$ Ω. The condition $R_I \sim 0.05$ Ω can be met by assuming the interfacial resistance of Cu and Pt; $r_{Cu/Pt} = R_I w_M w_N \sim 1.0$ fΩ·m$^2$ [90, 92]. However, $R_M'$ in equation (10) is very sensitive to the values of $R_I$ and $\delta$. If $R_I$ is slightly larger than $R_M$ (> 0.07 Ω), $R_M'$ becomes negative and this situation is not realistic. In addition, it is very difficult to unambiguously determine $R_I$ (especially, $\rho_I$ and $\lambda_I$) and $\delta$ with the spin absorption device. For this reason, we have not used equation (10) in our analyses.

## 7. Spin Hall effects in other materials

In this section, we discuss SHEs in two different materials. One is a weak ferromagnet, Pd$_{1-x}$Ni$_x$ alloy and the other is an oxidized material, IrO$_2$.

### 7.1. Spin Hall effect in weak ferromagnet

Recently it has been reported that the SHE occurs even in ferromagnets such as Py [93, 94]. Here we study the SHE in one of the weak ferromagnets, Pd$_{1-x}$Ni$_x$ alloy [95]. The Curie temperature $T_C$ of this alloy can be controlled by the Ni concentration. In the present case, we chose $x = 0.07, 0.08$, and $0.09$, and $T_C$ correspondingly changes from 16 K ($x = 0.07$) to 32 K ($x = 0.09$). This $T_C$ was determined from the AHE measurement of the Pd$_{1-x}$Ni$_x$ Hall bar, fabricated at the same time as the Pd$_{1-x}$Ni$_x$ SH device.

Figure 15 shows $R_{ISHE}$ of Pd$_{0.92}$Ni$_{0.08}$ measured from 15 K to 25 K ($T_C = 21$ K). A typical ISHE signal in nonmagnetic metals was observed far below and above $T_C$ where $R_{ISHE}$ is flat above the saturation field (~2000 Oe) of the Py wire. Only in the vicinity of $T_C$, however, a qualitatively different behavior was detected. $R_{ISHE}$ decreases or increases even above the saturation field of the Py wire. To better understand the observed $R_{ISHE}$, we plotted the amplitude of $R_{ISHE}$, $\Delta R_{ISHE} \equiv (R_{ISHE}(H_{max}) - R_{ISHE}(-H_{max}))/2$ as a function of temperature. A clear dip and peak can be seen in $\Delta R_{ISHE}$ as a function of $T$ (see figure 16($a$)). Such dip and peak near $T_C$ were also observed in different Pd$_{0.92}$Ni$_{0.08}$ devices and in different Ni concentrations. Thus, the anomalous behavior only in the vicinity of $T_C$ should be correlated to the magnetic phase transition of the PdNi wire.

Now we focus only on the anomalous parts $\delta\Delta R_{ISHE}$ at the different Ni concentrations. For this purpose, we subtracted the background signal $\Delta R_{ISHE}^0$, attributed to the skew scattering of Ni impurities in Pd, and extracted $\delta\Delta R_{ISHE}$ as a function of the reduced temperature $(T - T_C)/T_C$. As can be seen in figure 16($b$), although the three PdNi devices have different $T_C$ and $\Delta R_{ISHE}$, the anomalous part $\delta\Delta R_{ISHE}$ appears to be universal near $T_C$: the three curves almost scale onto one. In fact, a similar anomaly was also reported in the AHE of pure Ni more than 50 years ago [96, 97], but it is qualitatively different from the present case. In the AHE, the anomaly of the Hall resistivity appears only below $T_C$, while in the present case the anomalous part in the ISHE $\delta\Delta R_{ISHE}$ shows the dip and peak structure below and above $T_C$, respectively.

The anomalies observed in both the AHE and ISHE can be explained by Kondo's model [98, 99]. The Kondo's model was originally developed to explain the anomaly in the AHE appearing only be-low $T_C$ [98]. Recently Gu *et al.* have further modified the original Kondo's model to explain the anomaly in the ISHE below and above $T_C$ [99]. The details on this model should be referred to [95, 99]. Here, we explain intuitively the difference between the AHE and ISHE.

We assume the following three points: 1) the numbers of spin-up and spin-down electrons are the same since we consider the AHE and ISHE only in the vicinity of $T_C$; 2) the skew scattering events occur only at the Ni sites, which indicates that the skew scattering is more dominant in PdNi alloys than the intrinsic SHE in Pd; 3) only the on-site spin correlations at the Ni sites are taken into account (in other words, we neglect spin-spin correlations from the neighboring sites). In this situation, the scattering amplitudes of spin-up and spin-down electrons are proportional to $\chi_1 + \chi_2$ and $-\chi_1 + \chi_2$, where $\chi_1$ and $\chi_2$ are a first-order nonlinear susceptibility and a second-order nonlinear susceptibility, respectively. We note here that $\chi_2$ is one-order higher than $\chi_1$ but these two coefficients appear in the same order with respect to the SO coupling in the *s-d* Hamiltonian. In the AHE, the scattering directions of spin-up and spin-down electrons are opposite to each other. Thus, the detected signal is proportional to $\chi_1$, as illustrated in figure 17(*a*). Kondo already pointed out this fact more than 50 years ago and explained the anomaly in the AHE of pure Ni only below $T_C$ [98]. On the other hand, in the ISHE which is a relatively new concept in magnetism, the scattering directions are the same. That is why the detected signal is proportional to $\chi_2$, as illustrated in figure 17(*b*). As can be seen in figure 17(*b*), the temperature dependence of $\chi_2$ qualitatively reproduces the anomaly in the ISHE of PdNi near $T_C$. The second-order nonlinear susceptibility $\chi_2$ had been a hidden parameter in the Hall measurement for a long time because of the difference in the scattering directions. But the ISHE, that enables us to convert the pure spin current into the charge current, reveals that one can also detect $\chi_2$ electrically.

*7.2. Spin Hall effect in iridium oxide*

The SHE can be measured even in an oxidized material. Iridium dioxide $IrO_2$ is often used as an electrode in various device applications, ranging from non-volatile ferroelectric memories to electrochemical devices. The ISHE of $IrO_2$ was measured by means of the spin absorption method in the lateral spin valve structure, as shown in figure 18. Relatively large ISHE signals were observed even at $T = 300$ K both for polycrystalline and amorphous $IrO_2$ wires because of their high resistivities ($\rho$ = 200 μΩ·cm for polycrystalline $IrO_2$ and $\rho$ = 570 μΩ·cm for amorphous $IrO_2$). The spin diffusion length of $IrO_2$ is also determined from the NLSV measurement with an insertion of the $IrO_2$ wire and is about 4 nm for the polycrystalline $IrO_2$. The SH angle of polycrystalline $IrO_2$ obtained from the 3D analysis amounts to be 0.04, which is comparable to Pt [100].

**8. Conclusions and future prospects**

We have reviewed the spin absorption method in the lateral spin valve structure to evaluate the SH angle and the spin diffusion length of conductors with strong SO interactions, such as Cu-based alloys, 4*d* and 5*d* transition metals, weak ferromagnets, and oxides. The advantages of this method are that 1) both the SH angle and the spin diffusion length, which are important parameters in modern spintronic devices, can be evaluated on the same sample; 2) both DSHE and ISHE can be measured just by swapping the current and voltage probes. Although one may claim that the SH device is much more complicated than simple ferromagnet/nonmagnet bilayer films employed for spin-pumping and ST-FMR measurements, the physics discussed in the SH device is rather simple and does not rely on complex spectral analyses in which one has to take into account a variety of magnetotransport effects and complex interfacial problems.

Firstly, extrinsic SHEs in Cu-based alloys such as Ir-doped Cu and Bi-doped Cu have been measured with the spin absorption method. As for CuIr alloys, the SH resistivity exhibits a linear increase with the resistivity induced by the Ir impurities up to 12 atomic percent. This linear variation clearly shows that the skew scattering mechanism is dominant in this system. The SH angle of CuIr, defined from the slope of the SH resistivity vs resistivity due to Ir impurities, can be evaluated by using the 1D and 3D spin diffusion models and is about 0.02. In contrast, the linear regime for CuBi alloys is limited within the lower concentration of Bi because of its poor solubility in Cu. However, the SH angle of CuBi is about 10 times larger than that of CuIr. In addition, the 1D model underestimates the SH angle compared to the 3D model. This originates from a relatively long spin diffusion length of low concentration of Bi in Cu. When the spin diffusion length is longer than the thickness of CuBi, the spin current generated at the Py and Cu interface is not vertically absorbed into the CuBi wire but rather spreads along the wire axis direction. Such a spreading of spin accumulation is not taken into account in the 1D model, and thus the 3D model enables us to obtain the SH angle correctly.

To verify whether the SH angle and spin diffusion length obtained with the spin absorption method are reasonable, one needs to compare them with the values obtained by means of different experimental methods. For this purpose, we have performed two different experiments. One is WAL measurements to double-check the spin diffusion lengths of nonmagnetic metals. The other is AHE measurements of CuMnX terneary alloys to double-check the SH angles of Cu-based alloys. Most importantly in these measurements, target nonmagnetic metals are in no direct contact with ferromagnets. This is quite essential since the proximity effect between the ferromagnet and nonmagnet with strong SO interaction may significantly influence the estimation of the intrinsic SH angle and spin diffusion length of the nonmagnet. We measured WAL curves of several nonmagnetic metals and obtained their SO lengths by fitting the WAL curve with the Hikami-Larkin-Nagaoka formula for quasi-1D wire. Since the SO length is almost the same as the spin diffusion length, the spin diffusion length can be evaluated by measuring the magnetoresistance of simple nonmagnetic-metal wire very precisely. In fact, the evaluated spin diffusion lengths from the WAL measurements are quantitatively consistent with those obtained with the SH devices. We also performed the AHE measurements of CuMnIr and CuMnBi alloys to estimate the SH angles as an alternative method. This method was already studied by Fert *et al.* in 1981 for CuMnIr alloys but here we have also tried with CuMnBi alloys. The SH angles obtained from the AHE measurements are also quantitatively consistent with those determined with the SH devices. These facts clearly verify that the spin absorption method is a quite reliable method to evaluate the SH angle and spin diffusion length of nonmagnetic metals with strong SO interaction.

The SHE measurements have also been performed for 4*d* and 5*d* transition metals. In recent spintronic research, the 4*d* and 5*d* transition metals such as Pt and Ta are frequently used. Thus, it is of importance to obtain the SH angles and the spin diffusion lengths of those metals correctly. For example, as for Pt which is the most popular SHE material, the SH angle evaluated with the SH device is about 0.02, which is smaller than that obtained with other methods such as spin pumping and ST-FMR measurements. Correspondingly, the spin diffusion length of Pt is also significantly differ-ent, ranging from 10 nm to 1 nm. However, we have already demonstrated that reliable SH angles and spin diffusion lengths can be evaluated with the spin absorption method for Cu-based alloys. The overestimated SH angle and underestimated spin diffusion length in spin pumping and ST-FMR measurements might be due to some other effects such as the proximity effect between ferromagnet and 4$d$ (5$d$) transition metal. Thus, in future, it is important to understand the detailed mechanism why and how such the proximity effect influences the estimation of the SH angle and the spin diffusion length of 4$d$ and 5$d$ transition metals.

The SHE measurements with the spin absorption method have also been performed for other materials such as an oxide $IrO_2$ and a weak ferromagnet PdNi. Especially, in the latter case, anomalous behavior in the ISHE has been detected only in the vicinity of the Curie temperature. This anomaly can be explained by considering the temperature dependence of second-order magnetic susceptibility based on Kondo's model. The result also indicates that the pure spin current is quite sensitive to spin fluctuations near critical temperatures. The subject about how the pure spin current is affected by the spin fluctuations could be further tested, for example, with much more complicated systems such as spin glasses and frustrated magnetic materials.

**Acknowledgements**

We acknowledge helpful discussions with A Fert, S Maekawa, T Ziman, B Gu, T Kato, and X-F Jin. We would like to thank Y Iye and S Katsumoto for the use of the lithography facilities. This work was supported by Grant-in-Aid for Scientific Research (A) (23244071) from the Ministry of Education, Culture, Sports, Science and Technology, Japan (MEXT), Grant-in-Aid for Research Activity Start-up (22840012) from MEXT, Grant-in-Aid for Young Scientists (B) (24740217) from MEXT, and also by Foundation of Advanced Technology Institute.

**Figure captions**

**Figure 1.** Schematics of spin valve structures. (*a*) Current-perpendicular-to-plane GMR spin valve. (*b*) Lateral spin valve. In the latter case, by flowing a charge current from a ferromagnetic wire (F; injector) to a nonmagnetic wire (N), a pure spin current can be obtained only on the right side of N wire. The arrows indicate the directions of magnetizations in the F wires.

**Figure 2.** (*a*) Direct SHE (DSHE) and (*b*) inverse SHE (ISHE). The solid, broken, and dotted arrows indicate the directions of electric charge current, spin current, and the motions of spin-up and spin down electrons.

**Figure 3.** Illustrations of (*a*) skew scattering and (*b*) side jump near a potential center.

**Figure 4.** (*a*) Schematic of NLSV measurement using the lateral spin valve structure, consisting of two F wires bridged by an N wire. By passing $I_C$ from one of the F wires (injector) to the N wire, spin accumulation $\delta\mu$ is generated at the interface between the F and N interface as shown in (*b*). In this process a pure spin current $I_S$ flows in the right side of the N wire. $I_S$ decays within the spin diffusion length $\lambda_N$. (*c*) Electrochemical potentials $\mu$ in the N and two F wires, depending on the two magnetizations of the F injector and detector. The voltage difference $\Delta V_S$ between the parallel and antiparallel states can be detected at the detector.

**Figure 5.** SH device based on the lateral spin valve structure. (*a*) Schematic of the ISHE measurement. The ISHE in an SHE material deflects spin-up and spin-down electrons |*e*| denoted by spheres with arrows to the same side. Other arrows indicate the electron motion direction. The magnetic field is applied along the hard direction of the F wires ($H_\perp$). (*b*) Schematic of NLSV measurement with an insertion of the SHE material. Because of a strong SO interaction of the SHE material, $I_S$ is preferentially absorbed into the SHE material, but not all of $I_S$ is absorbed. The rest of $I_S$ still flows in the N channel and detected at the F detector. The magnetic field is applied along the easy direction of the F wires ($H_{//}$).

**Figure 6.** (*a*) Standard spin diffusion picture based on the Elliott-Yafet mechanism and (*b*) spin diffusion under WAL picture. $L_s$ and $L_{SO}$ are the spin diffusion length and the SO length, respectively.

**Figure 7.** Typical scanning electron microscopy images of (*a*) a SH device and (*b*) a Pt wire ($w = 100$ nm, $t = 20$ nm, $l = 1.9$ mm) for WAL measurement. The width $w_M$ and thickness $t_M$ of the SHE material are 250 nm and 20 nm, respectively.

**Figure 8.** (*a*) DSHE and (*b*) ISHE resistances of $Cu_{99.5}Bi_{0.5}$ (open square), $Cu_{97}Ir_3$ (open triangle) and pure Cu (closed circle) measured at $T = 10$ K. (*c*) A typical AMR signal of the Py wire showing the saturation of the magnetization above 2000 Oe for $H_\perp$.

**Figure 9.** NLSV signals measured at $T = 10$ K with a $Cu_{99.5}Bi_{0.5}$ middle wire (open square) and a $Cu_{97}Ir_3$ middle wire (open triangle) as well as without any M wire (closed circle). The arrows repre-

sent the magnetization directions of the two Py wires (see figure 7(*a*)). Note that, for the NLSV measurement, the magnetic field $H_{//}$ is aligned along the easy axis of the Py wires.

**Figure 10.** Spin diffusion lengths $\lambda_M$ of (*a*) CuIr and (*b*) CuBi alloys at 10 K as a function of $\rho_{imp}$. The closed and open symbols correspond, respectively, to the 3D and 1D analyses. The broken lines in the figures indicate the thickness (20 nm) of the CuIr and CuBi wires.

**Figure 11.** (*a*) SH resistivity of CuBi alloys as a function of $\rho_{imp}$. The closed and open symbols correspond, respectively, to the 3D and 1D analyses. The SH angles $\alpha_H^{3D}$ and $\alpha_H^{1D}$ correspond to the slopes of solid and broken lines, respectively. The inset shows the resistivity induced by Bi impurities $\rho_{imp}$ as a function of Bi concentration. (*b*) For comparison, the SH resistivity of CuIr alloys is plotted as in (*a*). The linear variation of the SH resistivity with $\rho_{imp}$ is not limited by the solubility of Ir up to concentrations as large as 12%. (*c*), (*d*) 3D mappings of the spin accumulation voltages for the (*c*) $Cu_{99.5}Bi_{0.5}$ and (*d*) $Cu_{97}Ir_3$ SH devices calculated with SpinFlow 3D [49]. Reprinted figure with permission from [31]. Copyright (2012) by the American Physical Society.

**Figure 12.** (*a*) WAL curves of Cu wires with different thicknesses ($t_{Cu}$ = 20, 30, and 80 nm) measured at $T$ = 4 K. The width of the Cu wires is 100 nm. The broken lines are the best fits of equation (8). The triangle in the figure corresponds to $B^*$ ($= h/(\pi e w L_{SO})$). The spin diffusion length of Cu $\lambda_{Cu}$ is obtained by using equation (9). (*b*) Diffusion coefficient $D$ dependence of $\lambda_{Cu}$ measured at $T$ = 4 K. The inset shows $l_e$ (left) and $\lambda_{Cu}$ (right) as a function of $t_{Cu}$. The dashed line is a guide to the eyes. (*c*) WAL curve of a long $Cu_{99.5}Bi_{0.5}$ wire measured at $T$ = 4 K. The width of the $Cu_{99.5}Bi_{0.5}$ wire is 120 nm. The broken line is the same meaning as in (*a*).

**Figure 13.** (*a*) ISHE resistances $R_{ISHE}$ measured at 10 K for various 4*d* and 5*d* transition metals (TMs). The widths of TMs are 250 nm for Ta and Mo, and 200 nm for Pt, Pd, and Au. The distance $L$ between the two Py wires is fixed to be 1 μm, which is slightly different from in [36]. The other characteristic parameters are shown in table I. (*b*) NLSV signals $R_S$ with and without TM wires measured at 10 K. (*c*), (*d*) WAL curves of (*c*) Pt and (*d*) Au wires. The width is 100 nm for both cases. The broken lines are the same meaning as in figure 12(*a*). The spin diffusion length $\lambda$ is obtained using equation (9).

**Figure 14.** SH angles of various 4*d* (circle) and 5*d* (square) TMs. The closed and open symbols show the SH angles measured with the SH devices and evaluated with the 3D model, and those theoretically calculated with the model shown in [35], respectively.

**Figure 15.** 3D plot of ISHE resistance $R_{ISHE}$ of $Pd_{0.92}Ni_{0.08}$. The width of $Pd_{0.92}Ni_{0.08}$ is 100 nm. The *z*-axis and the color scale show the amplitude of $R_{ISHE}$. The *x*-axis and *y*-axis are the magnetic field $H$ along the hard direction of the Py wire, and the temperature $T$, respectively. The triangles and stars near $T_C$ (= 21 K) indicate the positions of dips and peaks in $R_{ISHE}$, respectively. Adapted by permission from Macmillan Publishers Ltd: *Nat. Commun.* from [89], copyright (2012).

**Figure 16.** (*a*) $\Delta R_{ISHE}$ as a function of $T$ for different Ni concentrations. $\Delta R_{ISHE}^0$ is the ISHE signal due to the skew scattering at the Ni impurities and $\delta\Delta R_{ISHE}$ is the anomalous part appearing only near

$T_C$. (*b*) $\delta\Delta R_{ISHE}$ as a function of the reduced temperature $(T - T_C)/T_C$. The solid lines in (*b*) show the theoretically calculated $\chi_2$. Adapted by permission from Macmillan Publishers Ltd: *Nat. Commun.* from [89], copyright (2012).

**Figure 17.** Schematics of (*a*) AHE and (*b*) ISHE near $T_C$ in a weakly ferromagnetic metal. The incident current is the charge current $J_C$ in the AHE and the pure spin current $J_S$ in the ISHE. The fluctuation of the localized spins near $T_C$ is indicated by the longer arrows with shades. The interaction between the localized spins and conduction electron spins (the shorter arrows) is also indicated with the light yellow cloud. The skew scattering probabilities for spin-up and spin-down conduction electrons, resulting in anomalous Hall or inverse spin Hall voltage, are represented by the long curved arrows. $\chi_1$ and $\chi_2$ are defined in the main text.

**Figure 18.** ISHE resistances $R_{ISHE}$ of (*a*) polycrystalline $IrO_2$ and (*b*) amorphous $IrO_2$. Only in this case, the thickness $t_M$ of $IrO_2$ is 15 nm, the width $w_M$ is 170 nm or 100 nm, and silver instead of copper is used as an N wire to transport the pure spin current generated at the Py wire. Adapted by permission from Macmillan Publishers Ltd: *Nat. Commun.* from [94], copyright (2013).

# Tables

**Table I.** Characteristics of various SHE materials measured below 10 K. Some of the raw data were already shown in [30, 31, 33, 36, 55]

| SHE material | $\rho_{xx}$ or $\rho_{imp}$ ($\mu\Omega\cdot$cm) | $\alpha_H^{3D}$ | $\alpha_H^{1D}$ | $\lambda_M^{3D}$ (nm) | $\lambda_M^{1D}$ (nm) | $(\sqrt{3}/2)L_{SO}$ (nm) |
|---|---|---|---|---|---|---|
| Cu$_{99}$Ir$_1$ | 3.1 | 0.023($\pm$0.006) | 0.021($\pm$0.006) | 36($\pm$7) | 27($\pm$5) | |
| Cu$_{99.7}$Bi$_{0.3}$ | 3.2 | −0.26($\pm$0.11) | −0.11($\pm$0.04) | 86($\pm$17) | 53($\pm$8) | 66($\pm$4) |
| Cu$_{99.5}$Bi$_{0.5}$ | 5.1 | −0.24($\pm$0.09) | −0.12($\pm$0.04) | 45($\pm$14) | 32($\pm$9) | 37($\pm$3) |
| Cu$_{99.5}$Pb$_{0.5}$ | 5.4 | −0.13($\pm$0.03) | −0.07($\pm$0.02) | 53($\pm$15) | 36($\pm$7) | |
| Ag$_{99}$Bi$_1$ | 6.8 | −0.023($\pm$0.006) | −0.016($\pm$0.005) | 29($\pm$6) | 23($\pm$5) | |
| Nb | 90 | −0.013($\pm$0.003) | −0.009($\pm$0.002) | 6.8($\pm$0.3) | 5.9($\pm$0.3) | |
| Ta | 330 | −0.008($\pm$0.002) | −0.004($\pm$0.001) | 3.0($\pm$0.4) | 2.7($\pm$0.4) | |
| Mo | 35 | −0.012($\pm$0.003) | −0.008($\pm$0.002) | 10($\pm$2) | 8.6($\pm$1.3) | |
| Pd | 10 | 0.006($\pm$0.002) | 0.004($\pm$0.001) | 12($\pm$2) | 13($\pm$2) | |
| Pt | 10 | 0.024($\pm$0.006) | 0.021($\pm$0.005) | 10($\pm$2) | 11($\pm$2) | 10($\pm$2) |
| Au | 4.0 | 0.014($\pm$0.002) | 0.010($\pm$0.002) | 40($\pm$16) | 33($\pm$9) | 38($\pm$4) |

**Table II.** Typical physical quantities at $T = 10$ K of N (Cu; $w_N = t_N = 100$ nm) and F (Py; $w_F = 100$ nm, $t_F = 30$ nm) wires consisting of NLSV device.

| Parameters | $\rho_N$ ($\mu\Omega\cdot$cm) | $\lambda_N$ (nm) | $\rho_F$ ($\mu\Omega\cdot$cm) | $\lambda_F$ (nm) | $p_F$ | $\beta$ | $\gamma$ | $r_b^*$ (f$\Omega\cdot$m$^2$) | $g_{\uparrow\downarrow}$ ($\Omega^{-1}\cdot$m$^{-2}$) |
|---|---|---|---|---|---|---|---|---|---|
| values | 1.5 | 1300 | 19 | 5 | 0.23 | 0.31 | 0.31 | 0.5 | $1\times10^{15}$ |

(a) 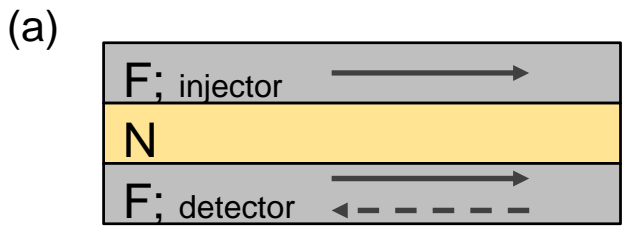

(b) 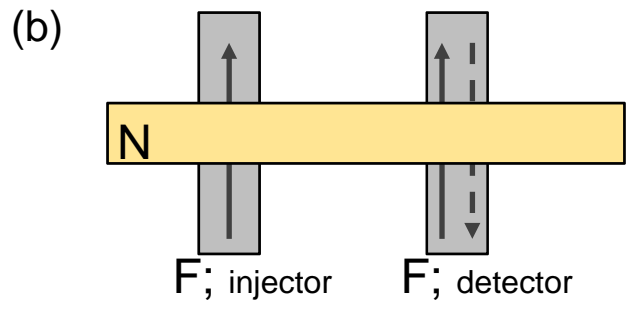

Figure 1.

(a) Direct spin Hall effect (DSHE)    (b) Inverse spin Hall effect (ISHE)

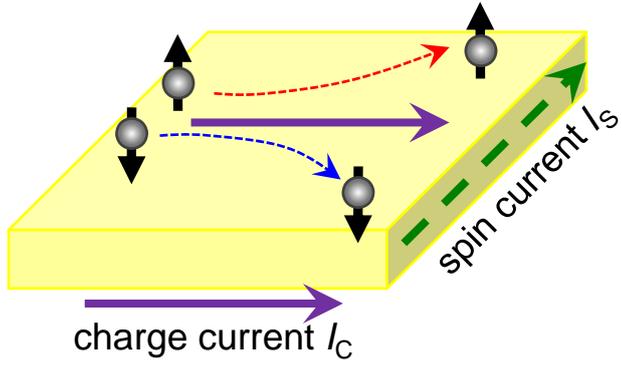 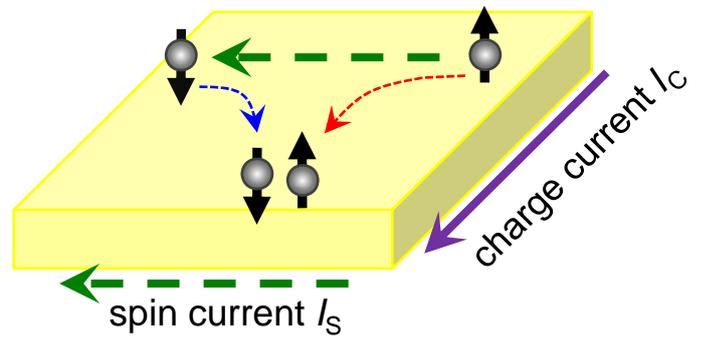

Figure 2.

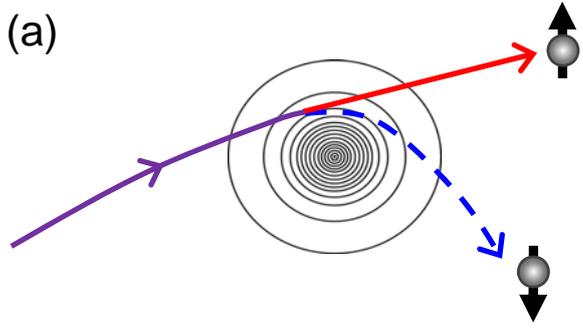
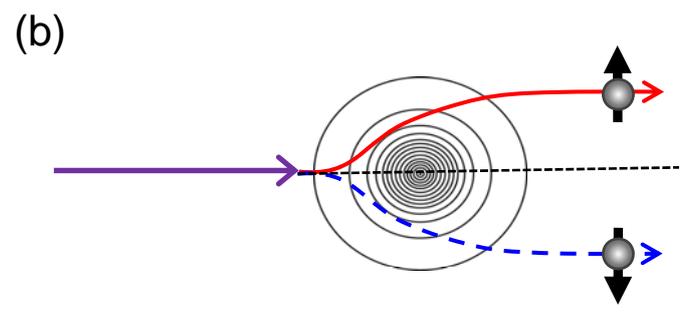

Figure 3.

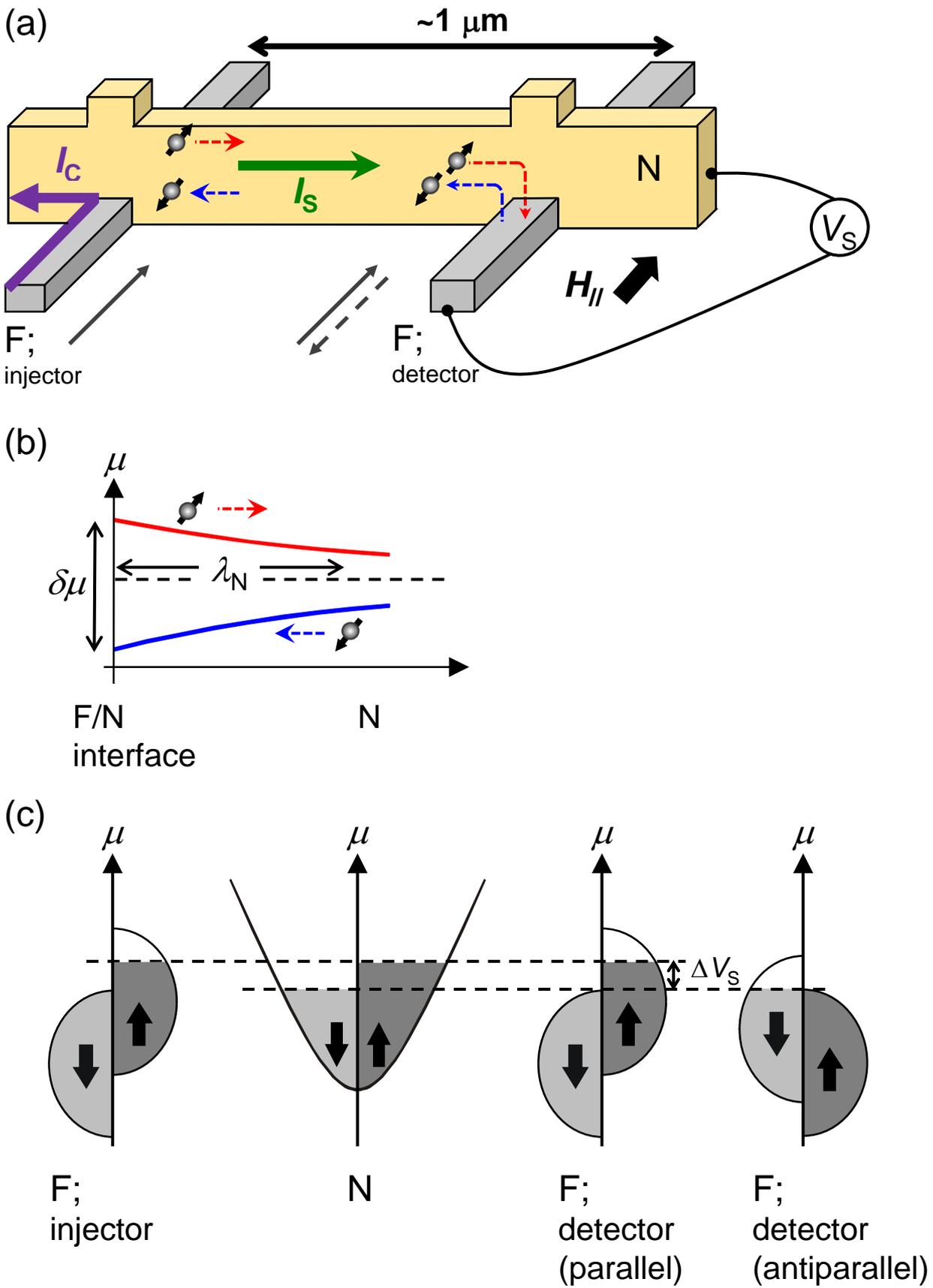

Figure 4.

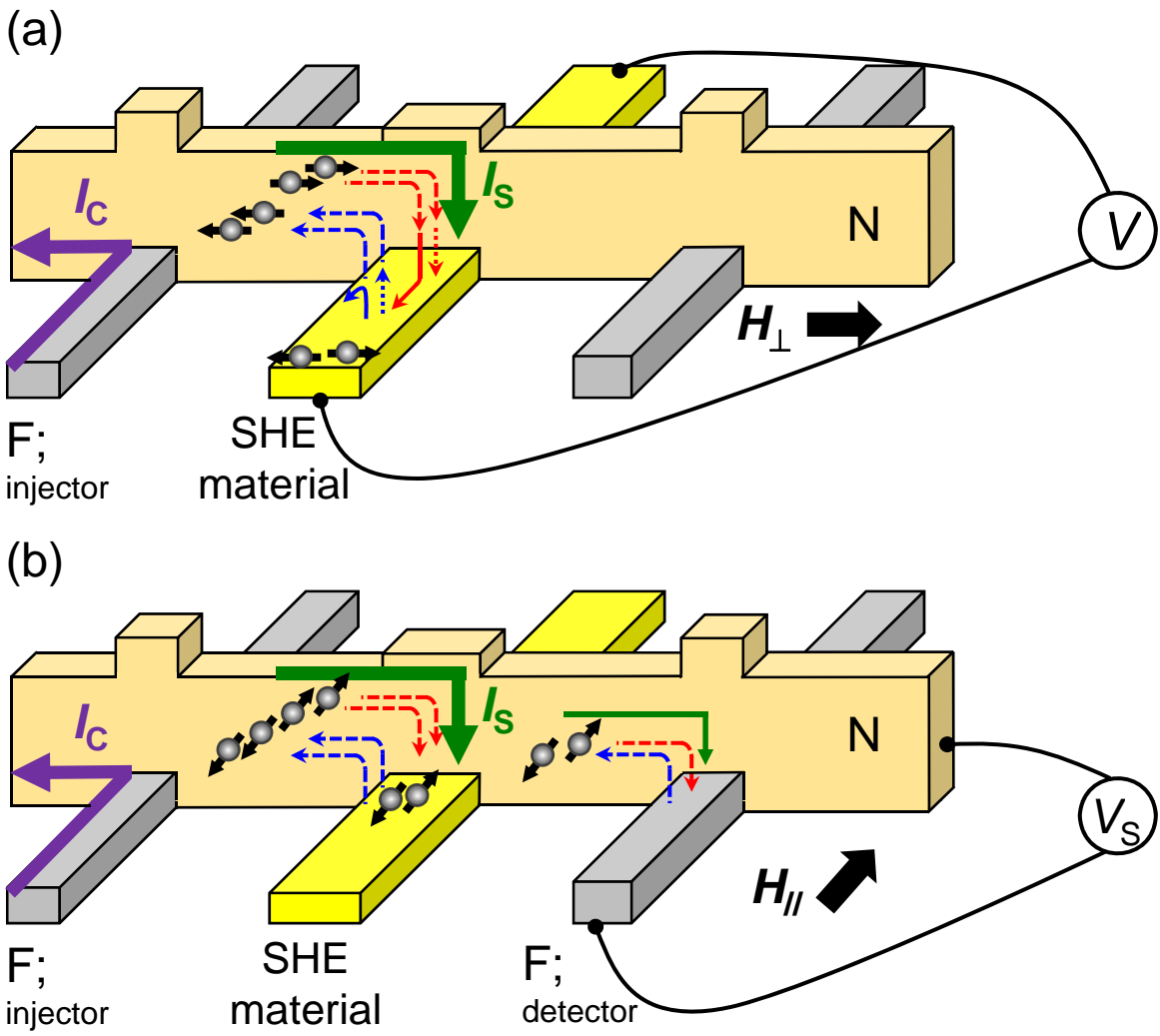

Figure 5.

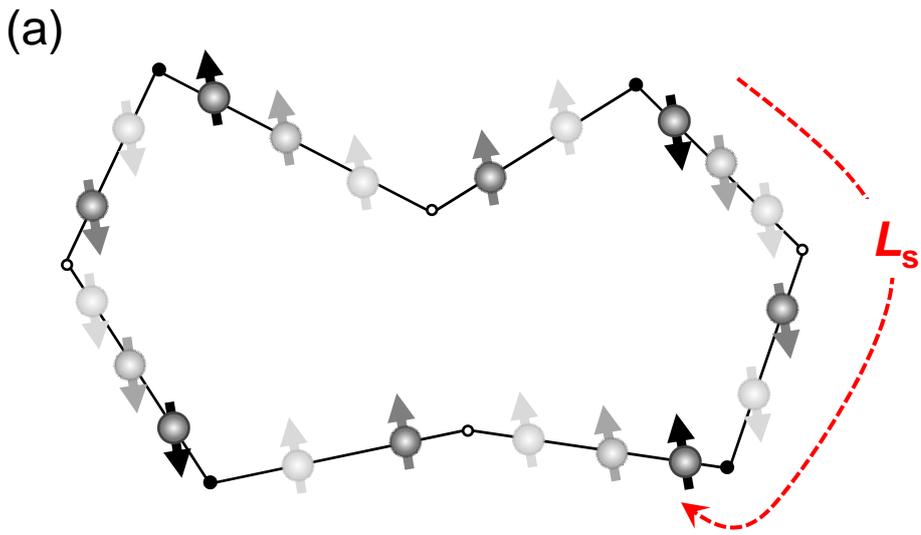

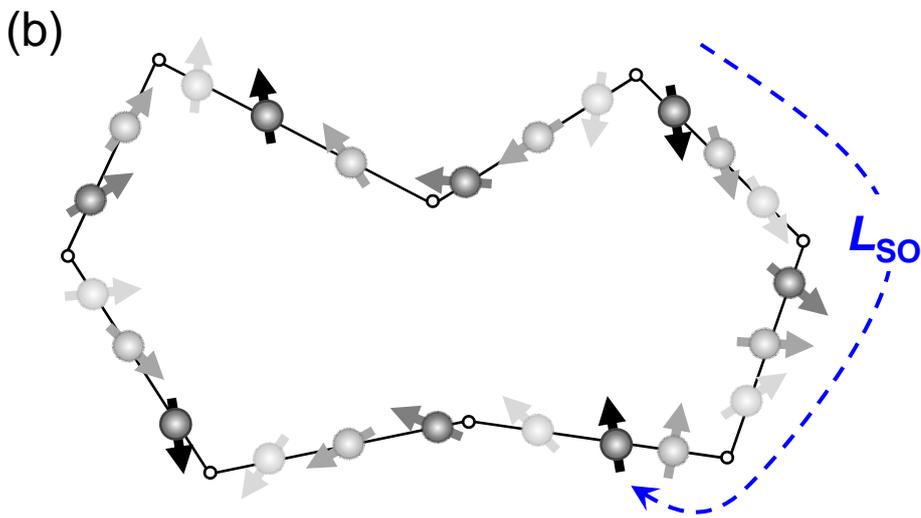

Figure 6.

(a) 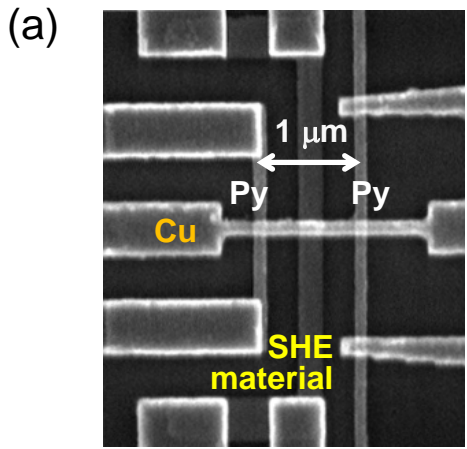 (b) 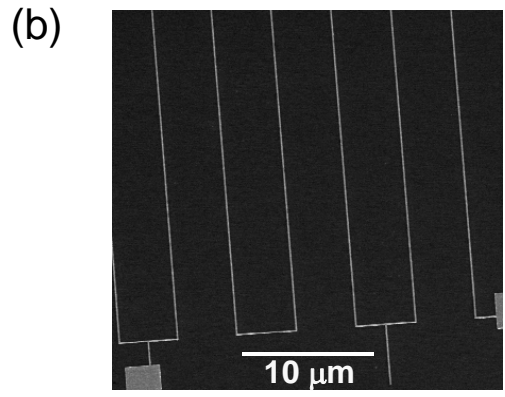

Figure 7.

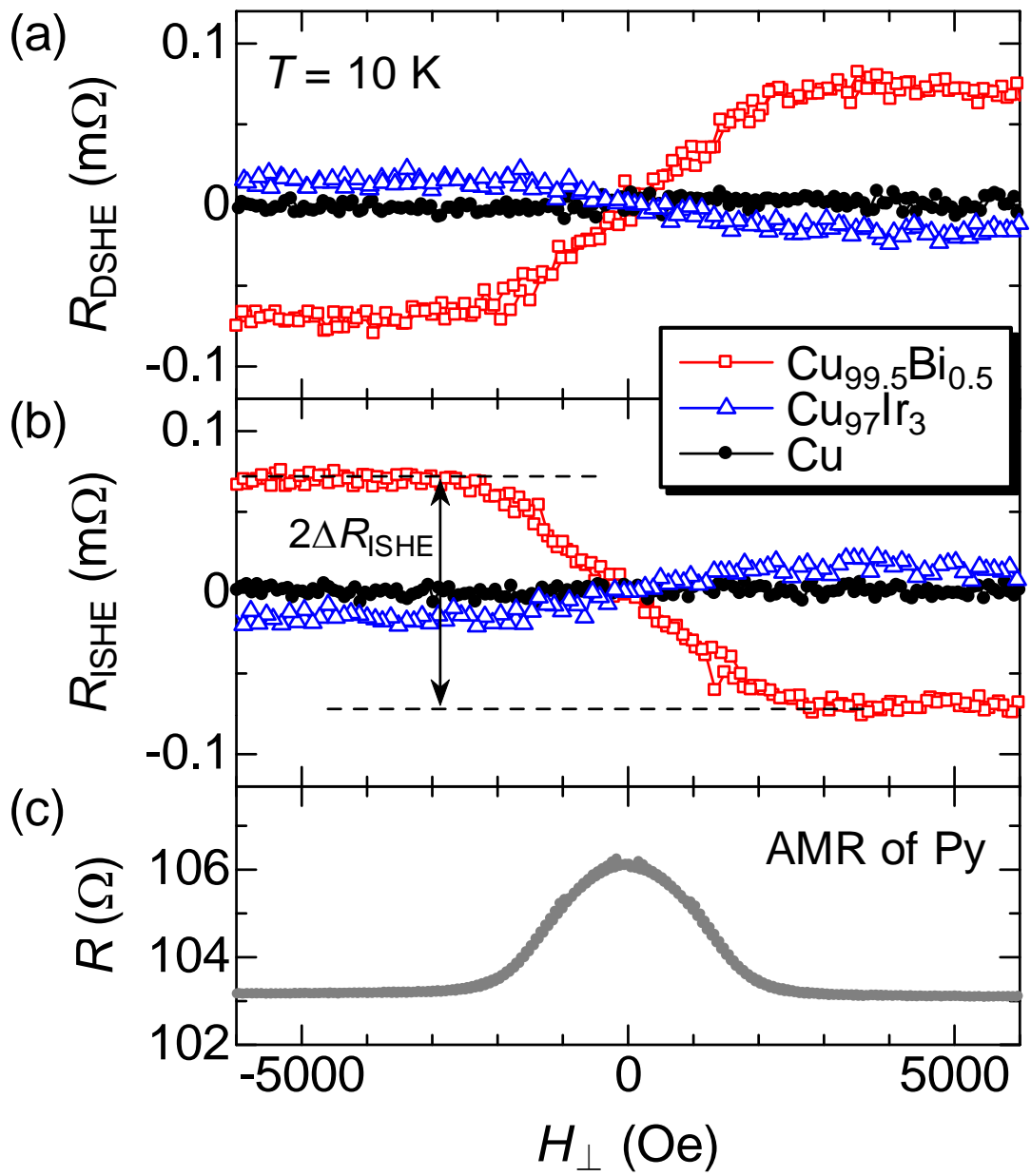

Figure 8.

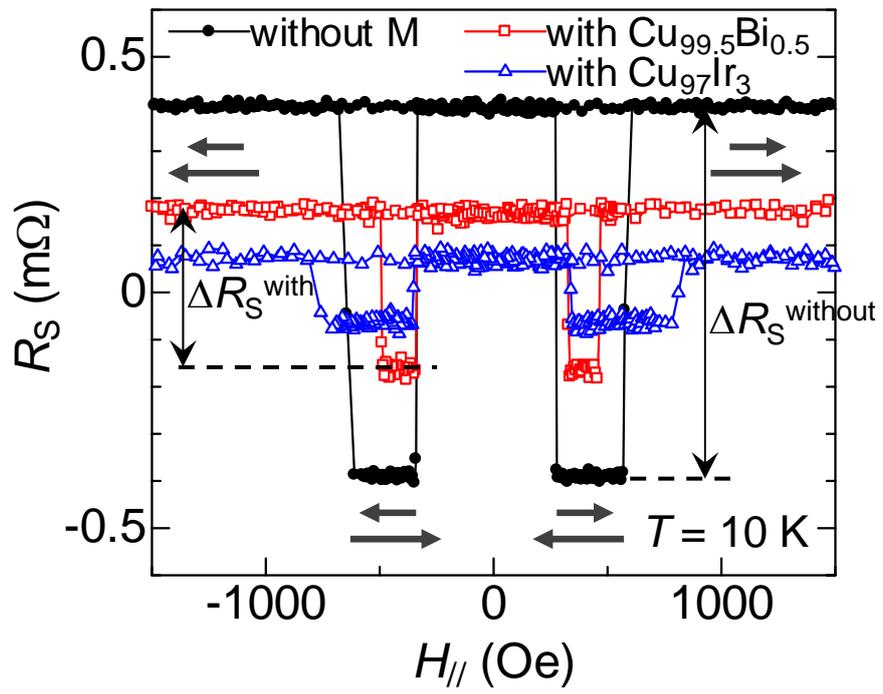

Figure 9.

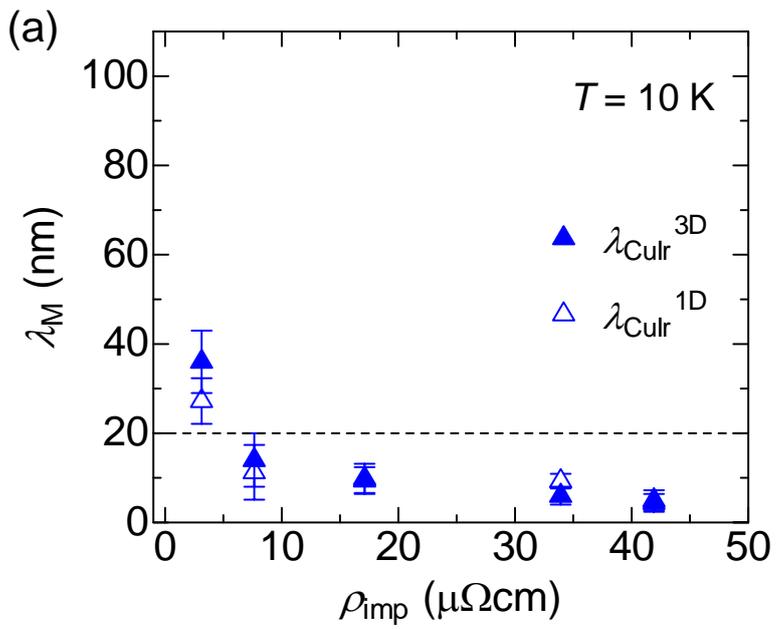
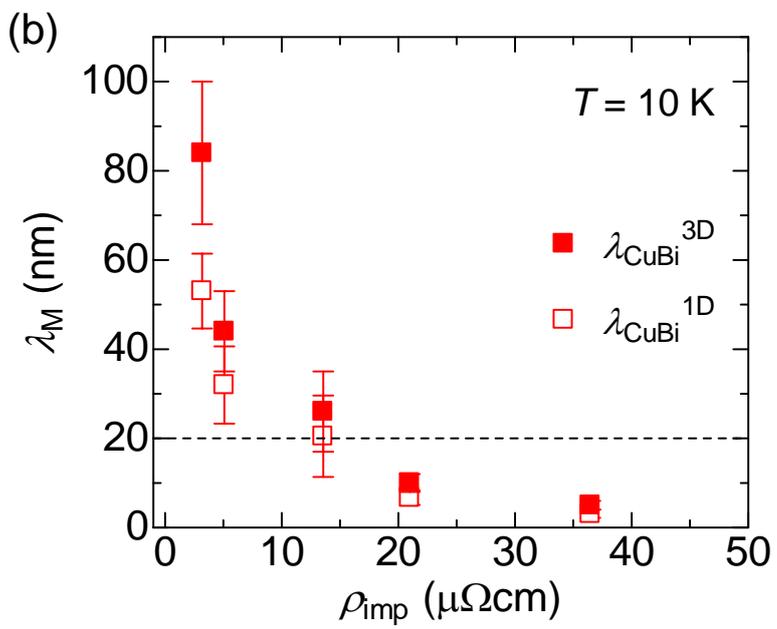

Figure 10.

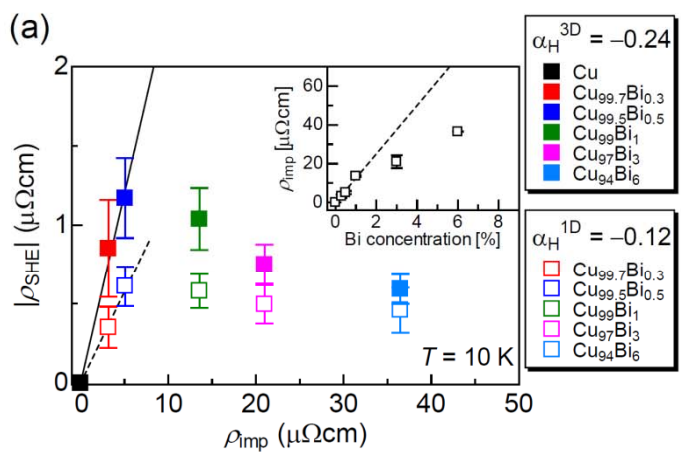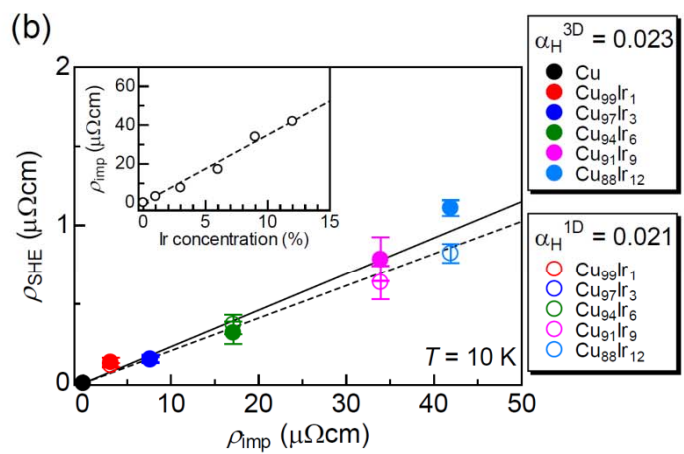
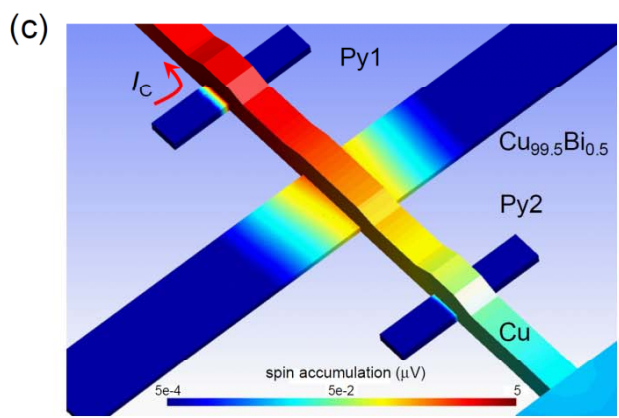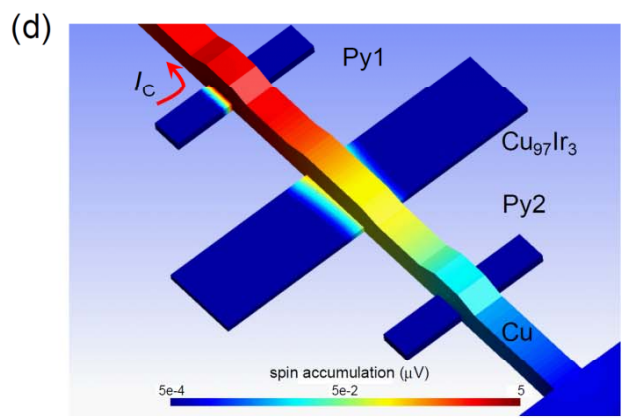

Figure 11.

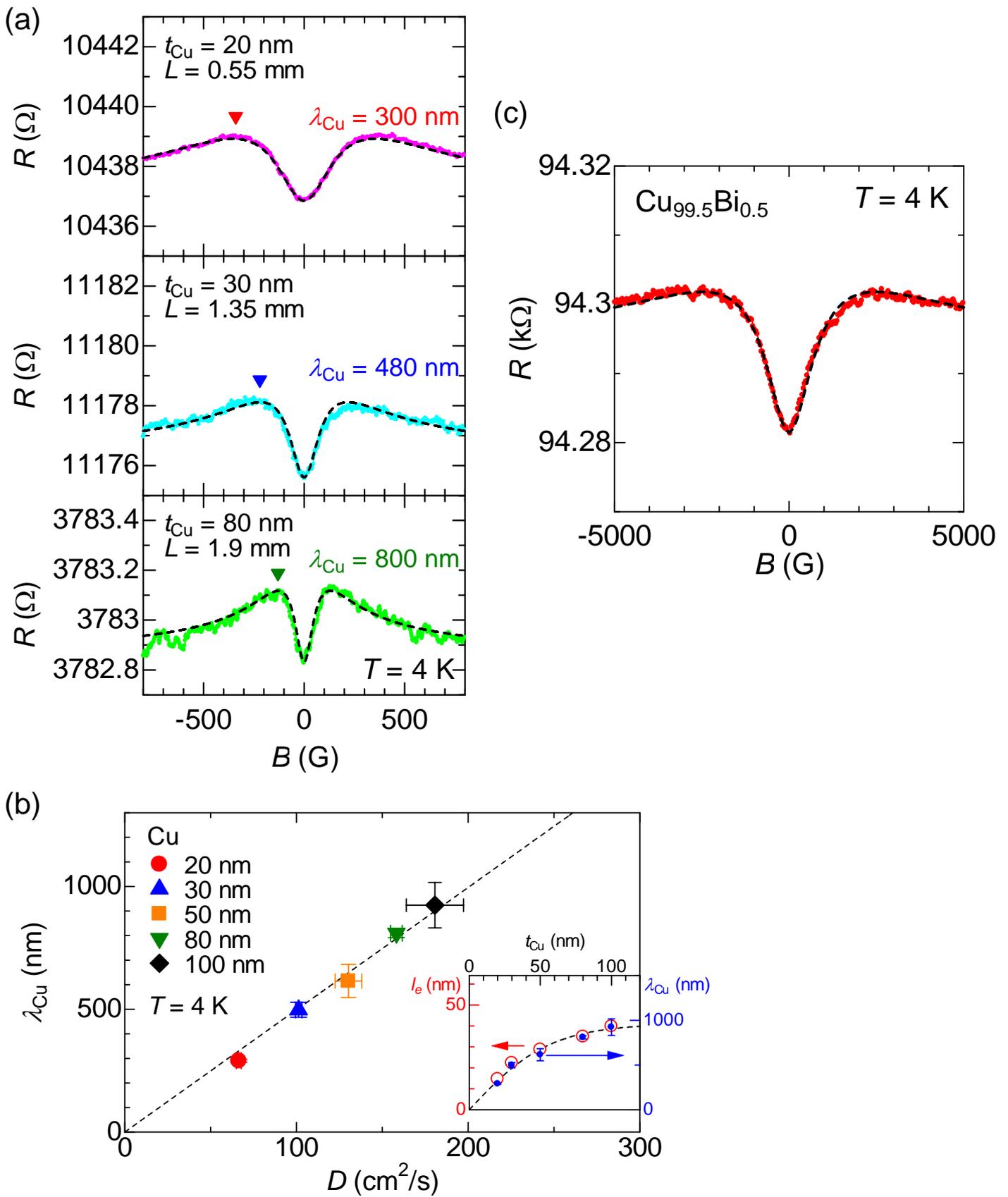

Figure 12.

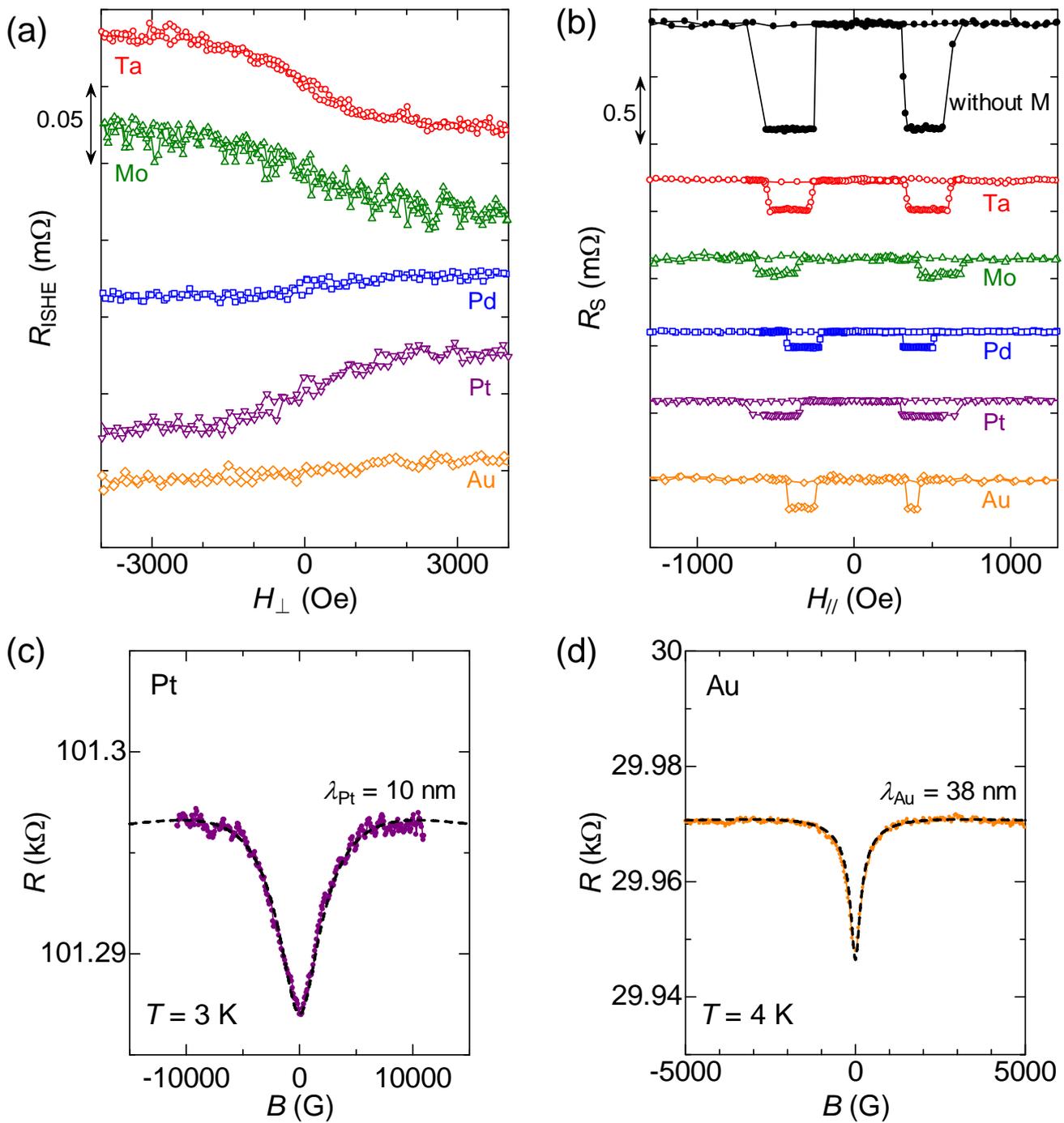

Figure 13.

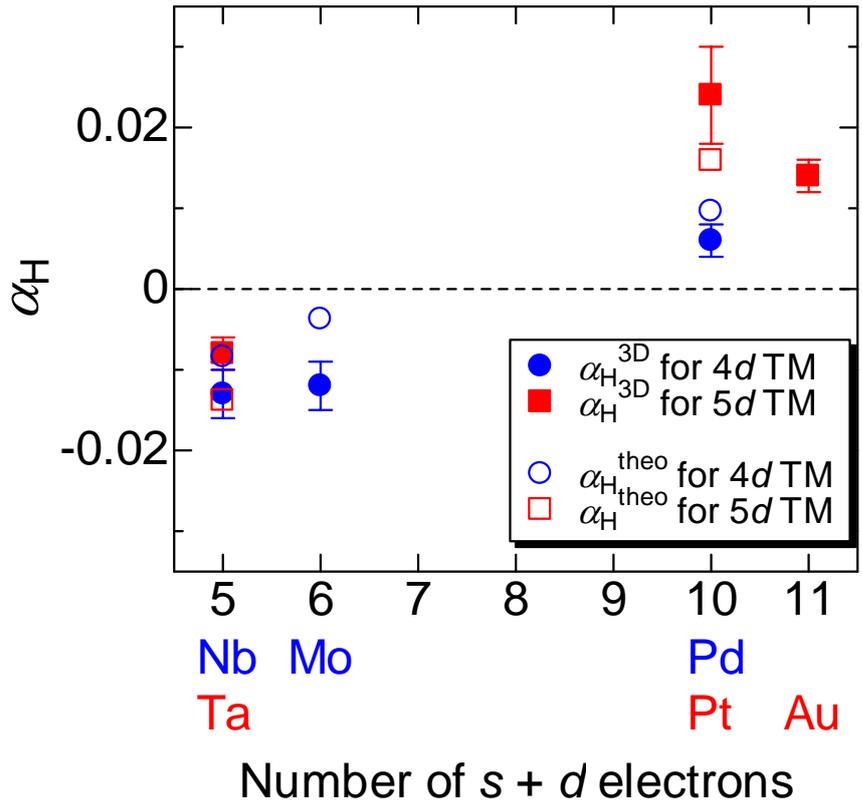

Figure 14.

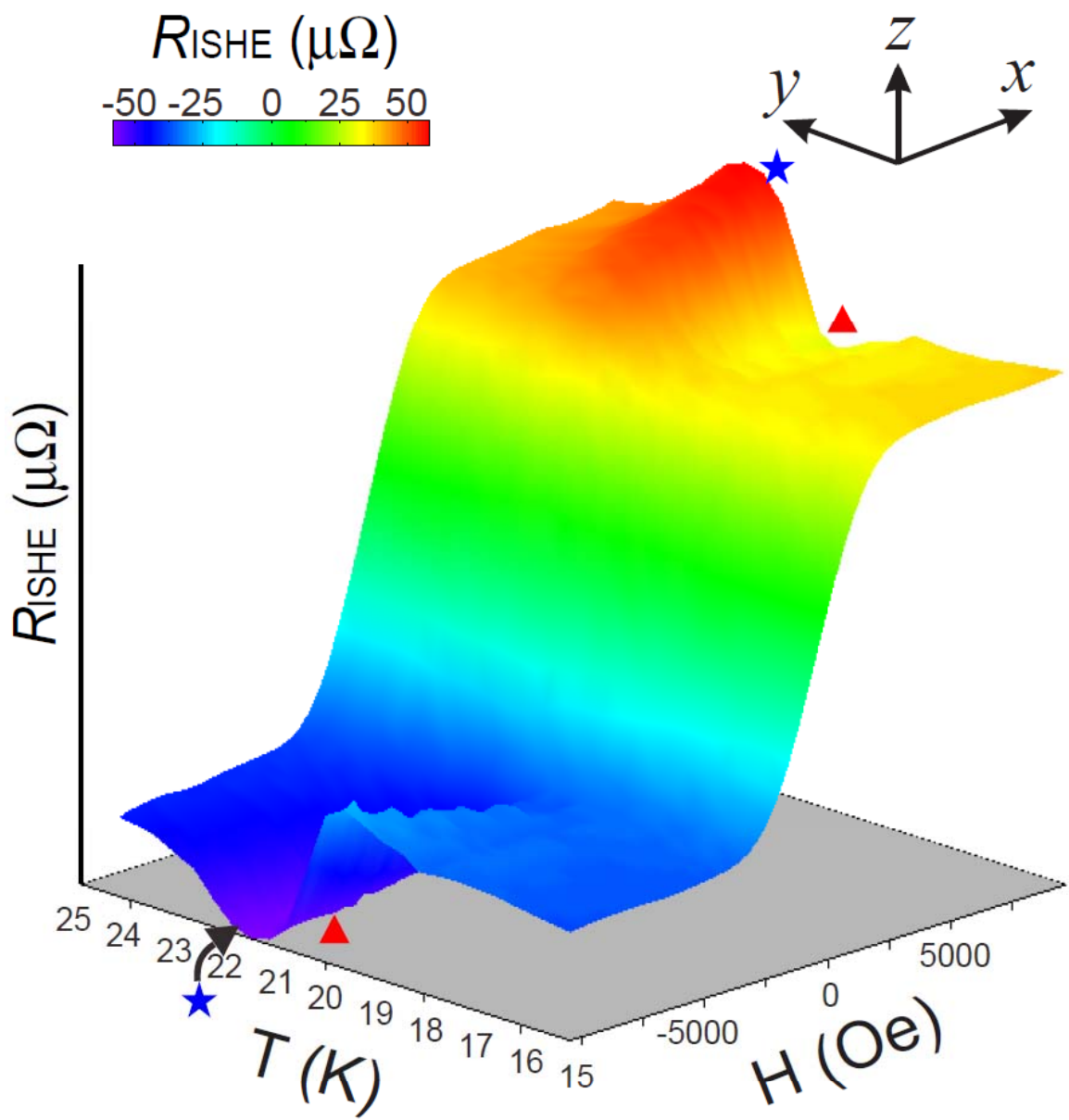

Figure 15.

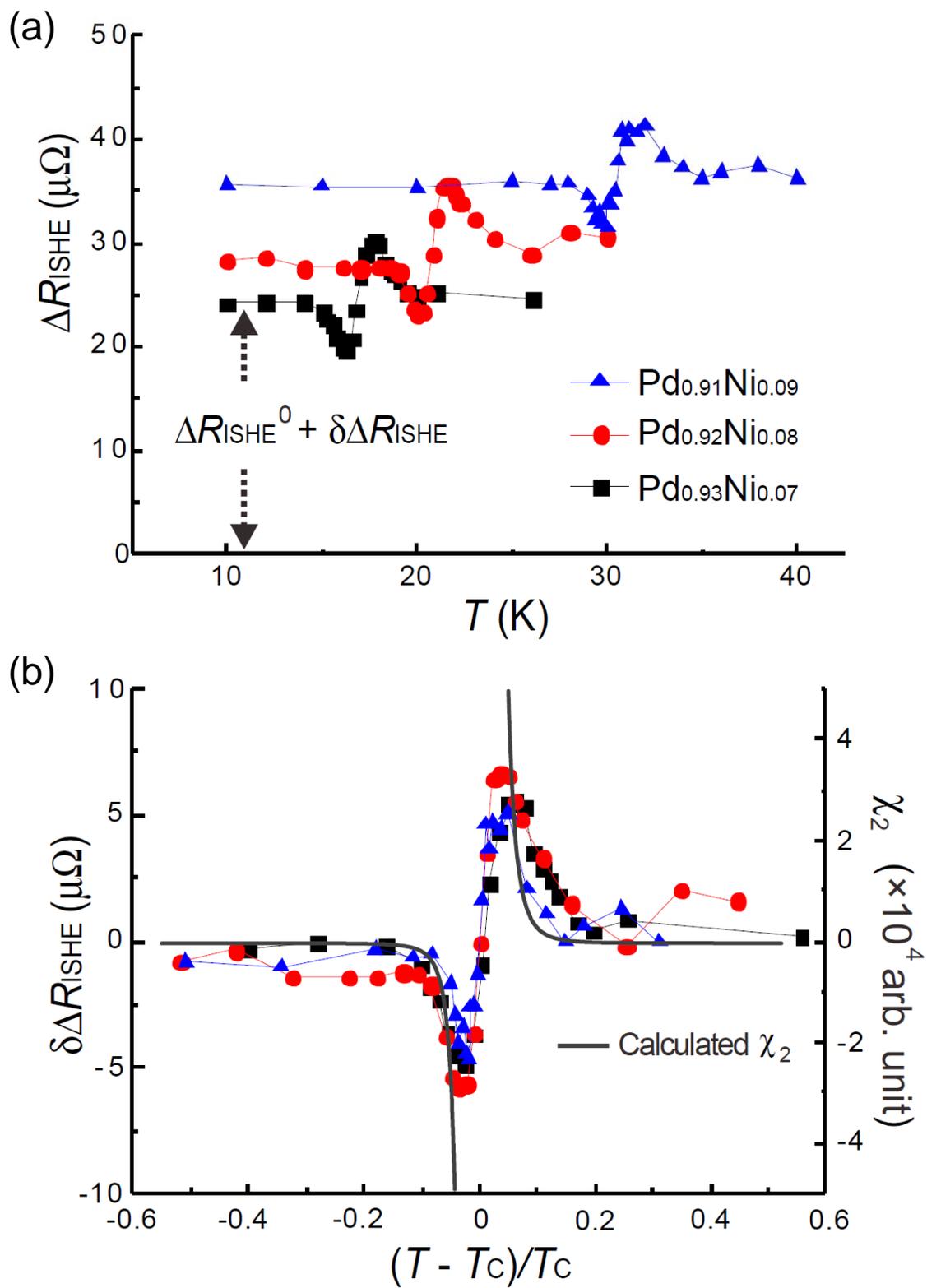

Figure 16.

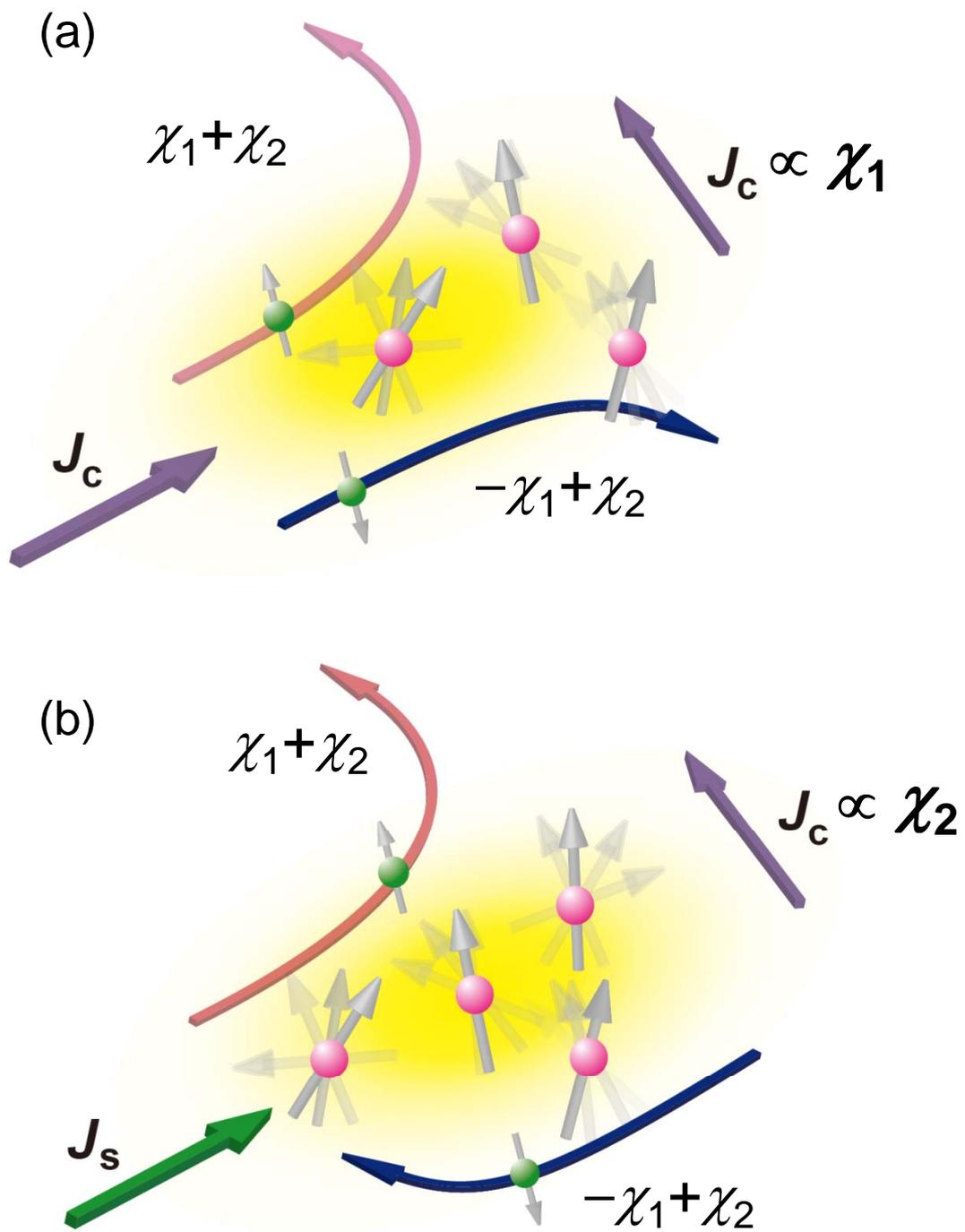

Figure 17.

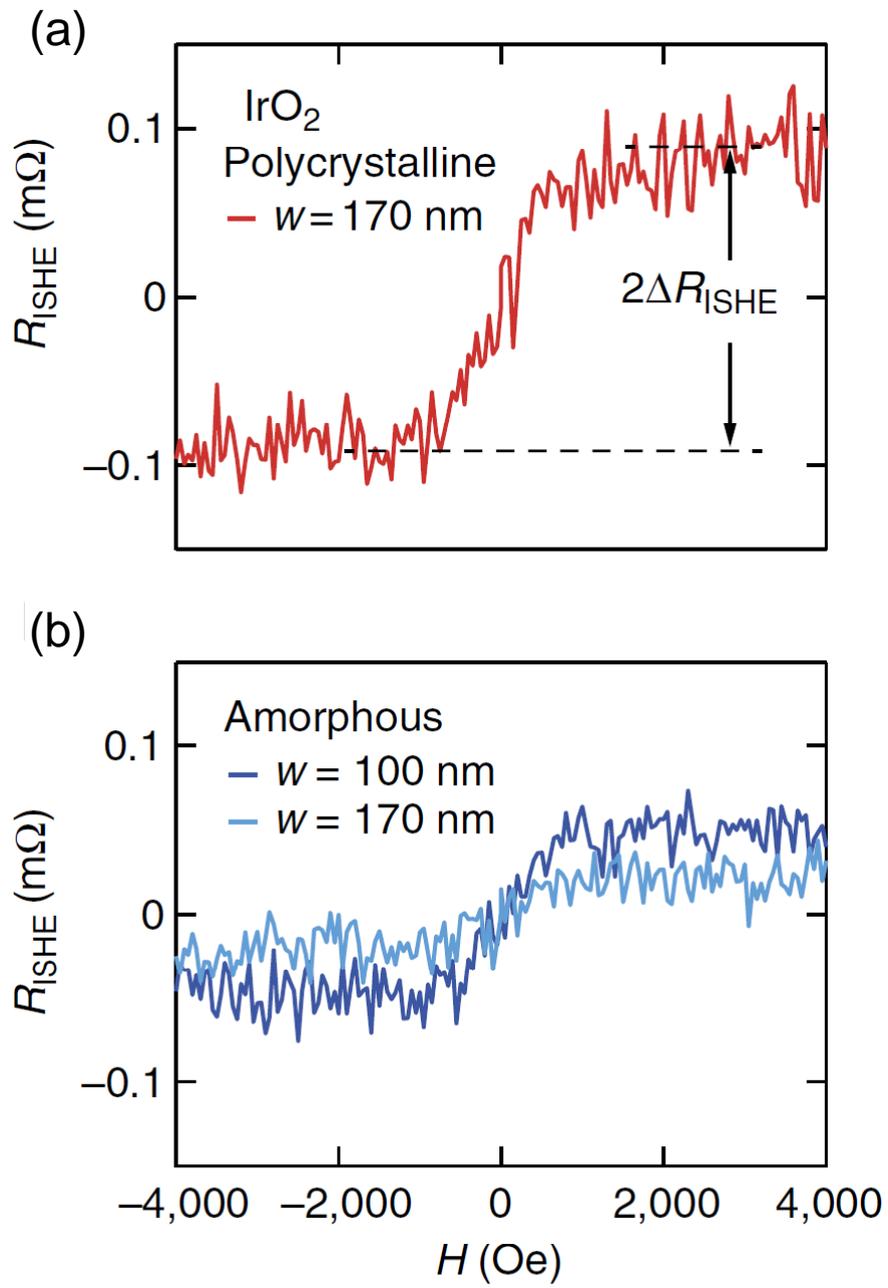

Figure 18.